\documentclass[12pt]{article}

\usepackage{amsmath,amsfonts}
\usepackage{amssymb}
\usepackage{hhline}
\usepackage{epsfig,cite}
\usepackage{stmaryrd}
\usepackage[usenames,dvips]{color}
\usepackage{fullpage}
\usepackage{verbatim}
   \allowdisplaybreaks
\makeatletter
\@addtoreset{equation}{section}
\makeatother

\newcommand{\be}{\begin{equation}}
\newcommand{\ee}{\end{equation}}
\newcommand{\bea}{\begin{eqnarray}}
\newcommand{\eea}{\end{eqnarray}}

\newcommand{\tr}{\operatorname{tr}}

\newcommand\T{\rule{0pt}{2.6pt}}

\begin{document}

\thispagestyle{empty}

\begin{center}
\hfill UAB-FT-742
\begin{center}

\vspace{.5cm}

{\Large\sc Supersymmetric Custodial Triplets}

\end{center}

\vspace{1.cm}

\textbf{ Luis Cort$^{\,a}$, Mateo Garcia$^{\,a}$,
and Mariano Quiros$^{\,b}$}\\

\vspace{1.cm}
${}^a\!\!$ {\em {Institut de F\'isica d'Altes Energies, Universitat Aut{\`o}noma de Barcelona\\
08193 Bellaterra, Barcelona, Spain}}

\vspace{.1cm}

${}^b\!\!$ {\em {Instituci\'o Catalana de Recerca i Estudis  
Avan\c{c}ats (ICREA) and\\ Institut de F\'isica d'Altes Energies, Universitat Aut{\`o}noma de Barcelona\\
08193 Bellaterra, Barcelona, Spain}}

\end{center}

\vspace{0.8cm}

\centerline{\bf Abstract}
\vspace{2 mm}
\begin{quote}\small
We analyze the extension of the Minimal Supersymmetric Standard Model which includes extra $Y=(0,\pm 1)$ supersymmetric triplets with a global $SU(2)_L\otimes SU(2)_R$ symmetry spontaneousy broken to the custodial $SU(2)_V$ by the vacuum expectation value of the neutral scalar components of doublets and triplets. The model is the supersymmetrization of the non-supersymmetric model introduced long ago by Georgi and Machacek where the $\rho$ parameter is kept to unity at the tree-level by the custodial symmetry. Accordingly the scalar sector is classified into degenerate $SU(2)_V$ multiplets: singlets, triplets (including the one containing the Godstone bosons) and fiveplets. The singly and doubly charged chiral superfields play a key role in the unitarization of the theory. The couplings of the Standard Model-like Higgs to vector bosons (including $\gamma\gamma$) and fermions, and the corresponding Higgs signal strengths, are in agreement with LHC experimental data for a large region of the parameter space. Breaking of custodial invariance by radiative corrections suggests a low-scale mechanism of supersymmetry breaking.

 \end{quote}

\vfill

 \newpage
 \section{Introduction}
 \label{introduction}
 Discovering the mechanism of electroweak symmetry breaking (EWSB) is one of the main theoretical issues in particle physics and one of the main experimental goals of the LHC. The recent discovery in the ATLAS~\cite{ATLAS:2013sla} and CMS~\cite{CMS:yva} collaborations of a resonance with a mass around 126 GeV, and with couplings to gauge bosons and fermions similar to those of the Standard Model (SM) Higgs, seems to point towards the Higgs doublet structure of the SM. However in view of possible departures of Higgs strengths with respect to the SM ones (as e.g.~in the $\gamma\gamma$ channel) it is interesting to search for possible extended Higgs structures and theories where other neutral and (singly or doubly) charged states do appear. In particular the inclusion of triplets is banned by precision observables: they contribute at tree level [when they get a vacuum expectation value (VEV)] in a non-custodial invariant way to the masses of the $W$ and $Z$ gauge bosons thus generating a large contribution to the $\rho$ parameter (i.e.~the $T$ parameter) which is forbidden by experimental data. Compatibility of scalar triplets with electroweak precision data was enforced by Georgi and Machacek (GM) in Ref.~\cite{Georgi:1985nv} where they introduced a global $SU(2)_R$ symmetry in the Lagrangian such that the electroweak vacuum respects the custodial symmetry $SU(2)_V$ subgroup of $SU(2)_L\otimes SU(2)_R$. This model was subsequently developed in a number of articles~\cite{Chanowitz:1985ug,Gunion:1989ci,Gunion:1990dt} and it has recently received a lot of attention in relation with the recent LHC results~\cite{GMstudies,1,2,3,4,5,6,7}.
 
 At a more theoretical level the SM Higgs square mass exhibits a quadratic sensitivity to the scale (a.k.a.~hierarchy problem) which makes it difficult to understand the big hierarchy between the electroweak scale and the (cutoff) scale below which we can believe the SM as an effective theory: Planck mass, unification scale, right-handed neutrino mass,~\dots This problem is shared by generic extensions of the SM as the two Higgs doublet model and of course the GM model~\cite{Georgi:1985nv} for which, on top of that, the $\rho$ parameter has been shown, see Ref.~\cite{Gunion:1990dt}, to exhibit a quadratic sensitivity to the scale which makes its value unpredictable. 
 
 In order to cope with these problem some extensions of the SM have been proposed. One of the most appealing and simple solutions is to introduce supersymmetry where the contribution to the Higgs square mass of the superpartners cancels in the radiative corrections the quadratic sensitivity induced by the SM particles. In particular the Minimal Supersymmetric extension of the SM (MSSM) was proposed as the simplest of such extensions where the hierarchy problem is solved. However the experimental value of the Higgs mass makes it necessary to introduce largish values of supersymmetric parameters in the  stop sector which generates a little hierarchy problem. In order to alleviate this problem the existing possibilities are: \textbf{i)} extending the gauge sector, and thus generating mass contribution from the extended $D$ term or, \textbf{ii)}  extending the Higgs sector, by introducing singlets and/or triplets and thus generating extra mass contributions from the extended $F$ terms. The latter possibility being the most economical one, as it does not require extending the gauge structure of the SM, has been considered since long ago by many authors~\cite{Espinosa:1991wt,uno,dos,tres,cuatro}, having in mind mainly the improvement on the fine-tuning as well as having extra contributions to the $\gamma\gamma$ rates~\cite{Agashe:2011ia,Basak:2012bd,Delgado:2012sm}. In particular extensions including Higgs triplets have the advantage of having extra singly and/or doubly charged states which contribute in loops to the $\gamma\gamma$ rates and can thus easily modify the SM (or the MSSM) decay rates in  case a diphoton excess be confirmed by the future LHC13-14. However extensions with triplets have the problem that the neutral triplet component acquire a VEV, which has to be small enough to cope with the present electroweak data, a problem of course shared with their non-supersymmetric partners. In order to implement a small enough VEV for the neutral triplet components, leaving apart an unnatural fine-tuning of supersymmetric parameters, the soft breaking mass of the triplet has to be in the TeV (or multi TeV) region where the scalar triplet essentially decouples from the MSSM Higgs sector. This in turn generates a little hierarchy problem as below the scale of triplet scalars the quadratic sensitivity produced by the triplet fermions is not canceled by the corresponding contribution from the triplet scalars. The only solution to this problem is to decouple the neutral triplet component VEV from the constrain from electroweak observables: the same problem that the GM model was intended to solve. In other words this problem can be solved by the supersymmetrization of the GM model.
 
 In this paper we provide a supersymmetric extension of the GM model and discuss some of its main theoretical and phenomenological features. In section~\ref{model} we extend the Higgs content of the MSSM by triplets with hypercharge $Y=0,\pm 1$ and define the superpotential and soft breaking terms which are invariant under the global $SU(2)_L\otimes SU(2)_R$. In particular the MSSM Higgs sector is classified into a bi-doublet and the extra triplets into a bi-triplet. We identify the custodially invariant minimum where the diagonal subgroup $SU(2)_V\subset SU(2)_L\otimes SU(2)_R$ is preserved and study the conditions for the EWSB mechanism. The lengthy and detailed expressions of the scalar potential in component fields is postponed to appendix~\ref{modelpotential} while some mathematical details on the way the vacuum preserves the custodial invariance are given in appendix~\ref{su2}. Finally the supersymmetric components of the MSSM doublets and triplets are decomposed into representations of $SU(2)_V$, i.e.~in singlets, triplets and fiveplets. In section~\ref{Higgssector} we analyze the Higgs sector of the present model and identify the mass eigenstates of the custodial invariant spectrum. In particular we find four singlets (two scalars and two pseudoscalars). The mass of one of the scalars ($S_1$) is not controlled by the supersymmetry breaking  (but by the electroweak breaking) scale and thus it can be identified with the SM-like Higgs. One of the pseudoscalars plays the role (in the limit when the neutral component of the triplets do not acquire any VEV) of the MSSM pseudoscalar. There are also four triplets, one of them is massless and contains the neutral and charged SM Goldstone bosons, and two fiveplets. The component fields of custodial multiplets are degenerate in mass, except for some tiny contribution $\mathcal O(g'^2)$ for the neutral with respect to the charged components of the triplets. No attempt has been made to scan over the large parameter space. Instead we have fixed the mass of the $S_1$ scalar to 126 GeV, chosen some generic values of the supersymmetric mass parameters and determined the supersymmetric coupling $\lambda$ of the bi-triplet to the MSSM Higgs sector as a function of the triplet VEV $v_\Delta$, $\lambda(v_\Delta)$. We have made a numerical analysis of the Higgs mass spectrum, as well as the location of the Landau pole, along the $\lambda(v_\Delta)$ trajectory. As for the location of the Landau pole, the renormalization group equations (RGE) as well as some notation are fixed in appendix~\ref{RGE}. Finally we have made some general considerations about the different decoupling regimes. In section~\ref{fermionsector} the fermion sector is considered in some detail, including seven neutralinos, and four singly charged and two doubly charged charginos. Numerical results for mass eigenvalues are presented along the $\lambda(v_\Delta)$ trajectory. 
 A short discussion on perturbative unitarity for the elastic scattering of longitudinal gauge bosons (in particular for $Z_LW_L\to Z_LW_L$ and $W_LW_L\to W_LW_L$) is performed in section~\ref{unitarity} where we can see the role played for the singly and doubly charged states for the unitarization of the theory. This calculation is also an interesting cross-check of the model couplings. The tree-level Higgs couplings to vector bosons and fermions are presented in section~\ref{Higgscouplings} in terms of the different mixing angles, and plots of the different couplings for the SM-like Higgs $S_1$ and its heavier orthogonal state $S_2$ along the $\lambda(v_\Delta)$ trajectory are shown. Likewise similar results for the one-loop coupling to $\gamma\gamma$ are presented in section~\ref{diphotonrate} and for the different signal strengths for the gluon-fusion and vector boson-fusion Higgs production mechanisms in section~\ref{gluonfusion}. 
 
 A discussion on the breaking of custodial symmetry by radiative corrections is done in section~\ref{breaking}. In particular we have assumed a set of supersymmetry breaking parameters respecting the superpotential global symmetry, i.e.~$SU(2)_L\otimes SU(2)_R$. Custodial symmetry at the electroweak scale is then exact \textit{only} if the mechanism of supersymmetry breaking respects the custodial symmetry group \textit{and} if the scale of supersymmetry breaking is around electroweak scale. Since we are assuming the former condition, but obviously not the latter, our setup is a good enough approximation only if supersymmetry breaking takes place at a low scale since the RGE evolution of supersymmetric masses will trigger departure from the custodial symmetry as the top quark Yukawa and hypercharge couplings break it explicitly. We have numerically analyzed this phenomenon by a small perturbation off the custodial minimum induced by the top Yukawa coupling (the leading effect breaking the custodial invariance). We have proved the self-consistency of the approximation as the relative departure of the minimum VEVs with respect to the exact custodial minimum is always below 10\%. Although the departure depends not only on the scale but also on the detailed mechanism of supersymmetry breaking we can set a reasonable absolute upper bound $v_\Delta\lesssim 25$ GeV. This value of $v_\Delta$ is enough to prevent the existence of superheavy scalar triplets (unlike in models with triplets and without custodial symmetry where phenomenological bounds on the $T$ parameter require scalars with masses $\gtrsim 1.5$ TeV~\cite{Delgado:2012sm}) as for our choice of parameters the heaviest scalar has a mass around 800 GeV. Finally section~\ref{conclusion} is devoted to our conclusions.

 \section{The model}
 \label{model}
 We will construct the Higgs sector manifestly invariant under $SU(2)_L\otimes SU(2)_R$. The MSSM Higgs sector $H_1$ and $H_2$ with respective hypercharges $Y=(-1/2,\,1/2)$ 
 \be
   H_1=\left( \begin{array}{c}H_1^0\\ H_1^-\end{array}\right),\quad
   H_2=\left( \begin{array}{c}H_2^+\\ H_2^0\end{array}\right)
   \ee
 is complemented with $SU(2)_L$ triplets, $\Sigma_{-1}$, $\Sigma_0$ and $\Sigma_1$ with hypercharges  $Y=(-1,\, 0,\, 1)$ 
 \be
 \Sigma_{-1}=\left(\begin{array}{cc} \frac{\chi^-}{\sqrt{2}} & \chi^0\\\chi^{--}& -\frac{\chi^-}{\sqrt{2}}
 \end{array}
 \right),\quad  \Sigma_{0}=\left(\begin{array}{cc} \frac{\phi^0}{\sqrt{2}} & \phi^+\\ \phi^{-}& -\frac{\phi^0}{\sqrt{2}}
 \end{array}
 \right),\quad  \Sigma_{1}=\left(\begin{array}{cc} \frac{\psi^+}{\sqrt{2}} & \psi^{++}\\\psi^{0}& -\frac{\psi^+}{\sqrt{2}}
 \end{array}
 \right)\ .
 \ee
where $Q=T_{3L}+Y$.

The two doublets and the three triplets are organized under $SU(2)_L\otimes SU(2)_R$ as $\bar H=(\textbf{2},\bar {\textbf{2}})$, and $\bar \Delta=(\textbf{3},\bar{\textbf{3}})$ where
\be
\bar H=\left( \begin{array}{c}H_1\\ H_2\end{array}\right),\quad
\bar\Delta=\left(\begin{array}{cc} -\frac{\Sigma_0}{\sqrt{2}} & -\Sigma_{-1}\\ -\Sigma_{1}& \frac{\Sigma_0}{\sqrt{2}}\end{array}\right)
\ee
and $T_{3R}=Y$.   The invariant products for doublets $A\cdot B\equiv A^a\epsilon_{ab}B^b$  and anti-doublets $\bar A\cdot \bar B\equiv\bar A_a\epsilon^{ab}\bar B_c$ are defined by $\epsilon_{21}=\epsilon^{12}=1$. 

The $SU(2)_L\otimes SU(2)_R$ invariant superpotential is then defined as
\be
W_0=\lambda \bar H\cdot \bar\Delta\bar H+\frac{\lambda_3}{3}\tr\bar\Delta^3+\frac{\mu}{2}\bar H\cdot\bar H+\frac{\mu_\Delta}{2}\tr \bar\Delta^2
\label{W0}
\ee
and the total potential
\be
V=V_F+V_D+V_{\rm soft}
\ee
where
\begin{eqnarray}
V_{\rm soft}&=&m_H^2|\bar H|^2+m_\Delta^2 \tr |\bar\Delta|^2+\frac{1}{2}m_3^2\bar H\cdot\bar H
\nonumber\\
&+&\left\{ \frac{1}{2}B_\Delta\tr\bar\Delta^2+A_\lambda \bar H\cdot \bar\Delta \bar H+\frac{1}{3}A_{\lambda_3}\tr\bar\Delta^3+h.c.\right\}
\label{Vsoft}
\end{eqnarray}
can be easily computed. The total expression in component fields can be found in appendix~\ref{modelpotential}.

The neutral components of all fields can be parametrized as
\be
X=v_X+\frac{X_R+i X_I}{\sqrt{2}},\quad X=H_1^0,H_2^0,\phi^0,\chi^0,\psi^0
\label{neutral}
\ee
 where we define by $v_1$, $v_2$, $v_\phi$, $v_\chi$ and $v_\psi$ the VEVs for the fields $H_{1R}^0$, $H_{2R}^0$, $\phi^0_R$, $\chi^0_R$ and $\psi^0_R$ respectively. When the fields VEVs are related by $v_1=v_2\equiv v_H$, $v_\phi=v_\chi=v_\psi\equiv v_\Delta$, the $SU(2)_L\otimes SU(2)_R$ symmetry is broken to the custodial (diagonal) subgroup $SU(2)_V$ (see appendix~\ref{su2}) and the (tree-level) parameter $\rho_0=1$. The tadpole conditions
 \be
 \left.\frac{\partial V}{\partial H_{1\,R}^0}\right|_0= \left.\frac{\partial V}{\partial H_{2\,R}^0}\right|_0
=\left. \frac{\partial V}{\partial \phi^0_R}\right|_0
= \left.\frac{\partial V}{\partial \chi^0_R}\right|_0
= \left.\frac{\partial V}{\partial \psi^0_R}\right|_0
=0
\label{minimo}
\ee
at the custodial VEV allow us to eliminate the parameters $m_H^2$ and $m_\Delta^2$  in the potential as a function of the other parameters as
\begin{align}
m_H^2&= m_3^2+3 v_\Delta[\lambda(\lambda_3 v_\Delta-\mu_\Delta)-A_\lambda]+6\lambda\mu v_\Delta-3\lambda^2(v_H^2+3 v_\Delta^2)-\mu^2
\nonumber\\
m_\Delta^2&=\frac{v_H^2(2\lambda\mu-6\lambda^2v_\Delta -A_\lambda)-v_\Delta B_\Delta+(2\lambda_3 v_\Delta-\mu_\Delta)[\lambda v_H^2-(\lambda_3 v_\Delta-\mu_\Delta)v_\Delta]-A_{\lambda_3}v_\Delta^2}{v_\Delta}
\label{condiciones}
\end{align}   

\begin{figure}[htb]
\begin{center}
\includegraphics[width=80mm]{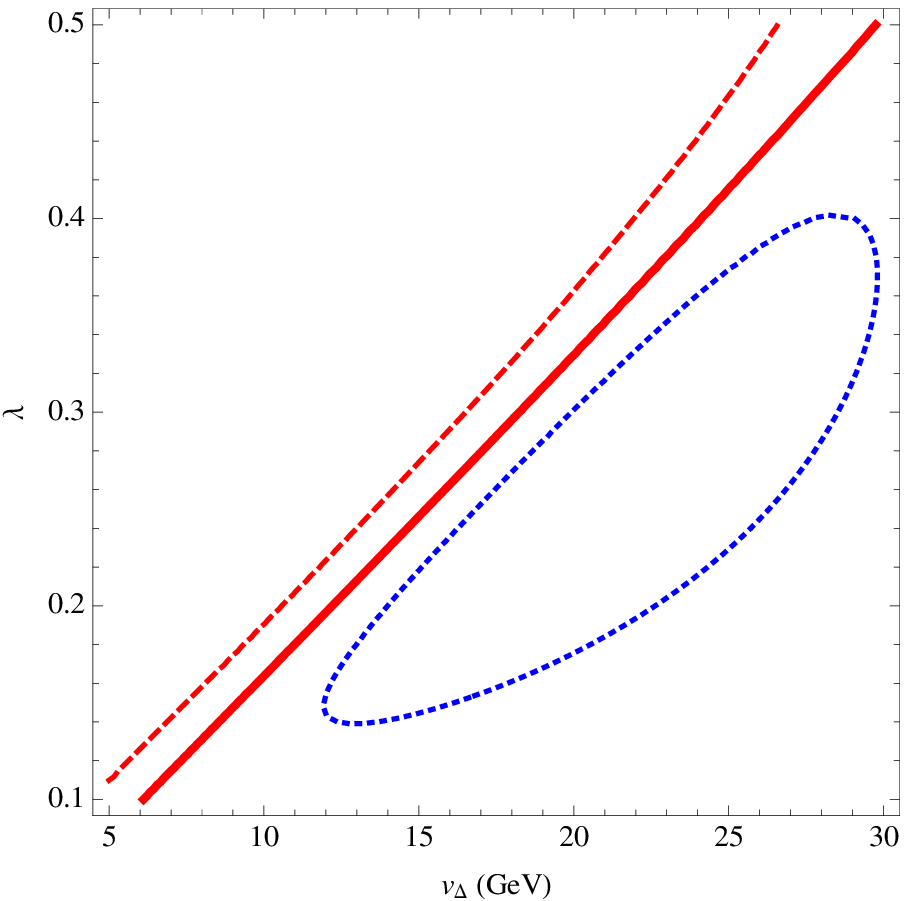}
\includegraphics[width=80mm]{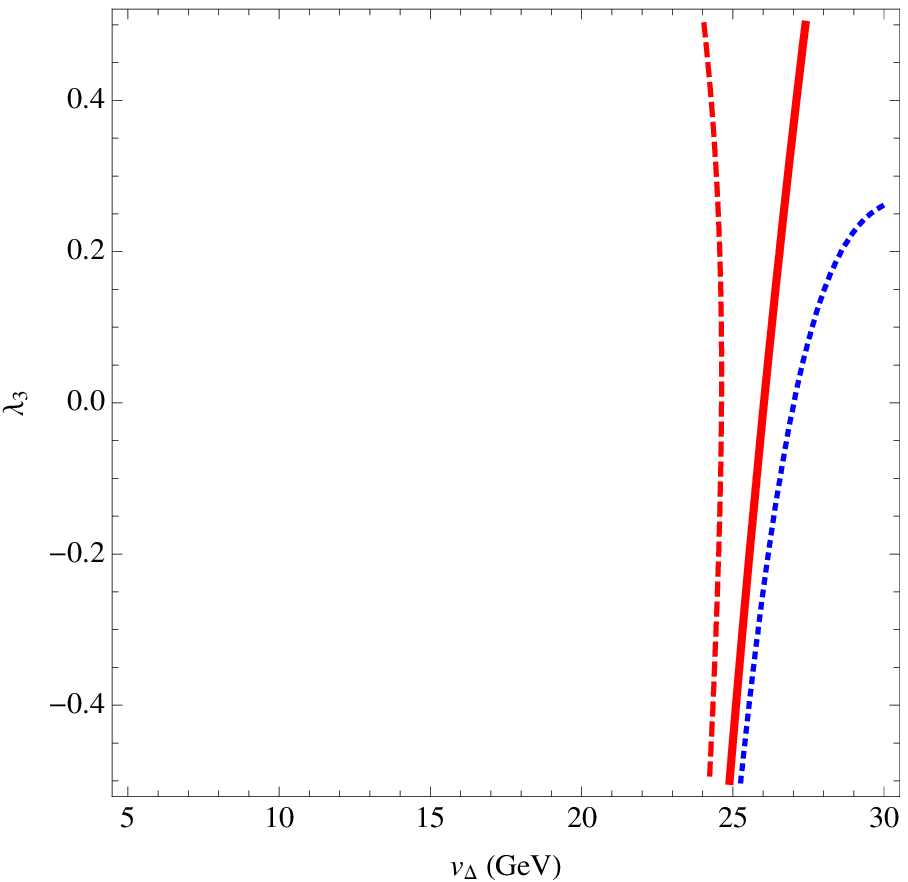}
\end{center}
\caption{\it Left panel: For the values of supersymmetric parameters given in Eq.~(\ref{valores}), contour lines of $\det H|_0/v^{10}=$-10 (dashed line), 0 (solid line) and 5 (dotted line) in the $(\lambda,v_\Delta)$ plane for $\lambda_3=-0.35$. Right panel: Contour lines of $\det H|_0/v^{10}=$-10 (dashed line), 0 (solid line) and 5 (dotted line) in the $(\lambda_3,v_\Delta)$ plane for $\lambda=0.45$.}
\label{ewbregion}
\end{figure}

Conditions (\ref{condiciones}) guarantee the existence of a non-trivial extremal but by no means enforce electroweak breaking. A set of sufficient conditions for the existence of a non-trivial minimum can be imposed by the condition $\det H|_0<0$ (where $H|_0$ is the Hessian matrix, or matrix of second derivatives, at the origin) which implies that the origin is a saddle point. This condition translates into a set of constraints in the space of supersymmetric parameters. To leading order in $v_\Delta$ these conditions can be written as
\begin{align}
\lambda(2\mu-\mu_\Delta)-A_\lambda>&0\nonumber\\
3 v_H^2\lambda^2-2 m_3^2<&0
\label{ewbconditions}
\end{align}
which hold for small values of $v_\Delta$. Of course when $v_\Delta$ grows conditions (\ref{ewbconditions}) are not a good approximation. In order to illustrate this fact we will consider here a particularly simple set of supersymmetric parameters 
\be
A_\lambda=A_{\lambda_3}=0,\quad \mu=\mu_\Delta=250\ \textrm{GeV},\quad m_3=500\ \textrm{GeV},\quad B_\Delta=-m_3^2
\label{valores}
\ee
which we will be using hereafter. We show in Fig.~\ref{ewbregion} the plot of $\det H|_0/v^{10}$, where $v^2=2 v_H^2+8 v_\Delta^2$,  in the $(\lambda,v_\Delta)$ plane for $\lambda_3=-0.35$ (left panel) and in the $(\lambda_3,v_\Delta)$ plane for $\lambda=0.45$ (right panel).
In both plots the region on the left (right) of the thick solid line fulfills (does not fulfill) the electroweak breaking condition. In the plots of Fig.~\ref{ewbregion} we can see that, for fixed values of the supersymmetric parameters, there is an upper bound on the value of $v_\Delta$ such that beyond the bound electroweak symmetry breaking does not hold. As we will see in the next sections the chosen values of supersymmetric parameters are consistent with a SM-like Higgs with a mass $\sim 126$ GeV.

Before discussing the mass spectrum we will change the states $\bar H$ and $\bar\Delta$, which are representations of the Lagrangian group $SU(2)_L\otimes SU(2)_R$ symmetry, into representations of the custodial vacuum $SU(2)_V$ symmetry. To this end we will decompose the representations as $\bar H=h_1\oplus h_3$ and $\bar\Delta=\delta_1\oplus\delta_3\oplus\delta_5$ where the subscripts indicate the dimensionality of the $SU(2)_V$ representations and
\begin{align}
h_1^0&=\frac{1}{\sqrt{2}}(H_1^0+H_2^0)\nonumber\\
h_3^+&=H_2^+,\quad
h_3^0=\frac{1}{\sqrt{2}}(H_1^0-H_2^0),\quad
h_3^-=H_1^-
\label{multH}
\end{align}
 and
 \begin{align}
 \delta_1^0&=\frac{\phi^0+\chi^0+\psi^0}{\sqrt{3}}\nonumber\\
 \delta_3^+&=\frac{\psi^+-\phi^+}{\sqrt{2}},\ \delta_3^0=\frac{\chi^0-\psi^0}{\sqrt{2}},\
 \delta_3^-=\frac{\phi^--\chi^-}{\sqrt{2}} \label{multD}\\
 \delta_5^{++}&=\psi^{++},\, \delta_5^+=\frac{\phi^++\psi^+}{\sqrt{2}},\, \delta_5^0=\frac{-2\phi^0+\psi^0+\chi^0}{\sqrt{6}}
 ,\, \delta_5^-=\frac{\phi^-+\chi^-}{\sqrt{2}},\, \delta^{--}_5=\chi^{--}\nonumber
  \end{align}
 Notice that the field components of $h_{1,3}$ and $\delta_{1,3,5}$ are complex. After electroweak breaking they decompose into real representations of $SU(2)_V$ with a common mass for all components, including the massless Goldstone triplet. We then decompose the neutral components of fields in (\ref{multH}) and (\ref{multD}) as
 \begin{align}
 h_1^0&=\sqrt{2}v_H+\frac{h^0_{1R}+i h_{1I}^0}{\sqrt{2}}\nonumber\\
 \delta_1^0&=\sqrt{3}v_\Delta+\frac{\delta^0_{1R}+i \delta_{1I}^0}{\sqrt{2}}\nonumber\\
 h_3^a&=\frac{h_{3R}^a+i h_{3I}^a}{\sqrt{2}},\quad \delta_3^a=\frac{\delta_{3R}^a+i \delta_{3I}^a}{\sqrt{2}}\ (a=+,0,-)\nonumber\\
 \delta_5^A&=\frac{\delta_{5R}^A+i \delta_{5I}^A}{\sqrt{2}}\ (A=++,+,0,-,--)
\label{neutralmultiplets}
\end{align} 

 \section{The Higgs sector}
 \label{Higgssector}
 We will describe in this section the spectrum of the scalar and pseudoscalar sectors~\footnote{By an abuse of language we will sometimes refer to a scalar (pseudoscalar) multiplet as one whose neutral component is a scalar (pseudoscalar).} after electroweak breaking in the custodial minimum. Because of the residual custodial invariance of the Higgs sector the mass eigenstates transform as representations of the custodial group $SU(2)_V$ (i.e.~singlets, triplets and fiveplets) which makes it possible to compute analytical expressions for the mass eigenvalues and mixing angles. 

\subsection{The $SU(2)_V$ triplet sector} 
We will first describe the triplet sector which in particular contains the Goldstone triplet. There are two pseudoscalar triplets, the massless triplet $G=\left(G^+,G^0,G^-\right)^T$ describing the massless Goldstone bosons and the massive triplet $A=\left(A^+,A^0,A^-\right)^T$, with components
   \begin{align}
   G^0&=\cos\theta\, h_{3I}^0+\sin\theta\, \delta_{3I}^0\nonumber\\
G^\mp&=\cos\theta\,\frac{h_3^{\pm *}-h_3^\mp}{\sqrt{2}}+\sin\theta\, \frac{\delta_3^{\pm *}-\delta_3^\mp}{\sqrt{2}}\nonumber\\
   A^0&=-\sin\theta\, h_{3I}^0+\cos\theta\, \delta_{3I}^0\nonumber\\
  A^\mp&=-\sin\theta\,\frac{h_3^{\pm *}-h_3^\mp}{\sqrt{2}}+\cos\theta\, \frac{\delta_3^{\pm *}-\delta_3^\mp}{\sqrt{2}}
   \end{align}
  where the mixing angle is defined as
  \be
  \sin\theta=\frac{2\sqrt{2}v_\Delta}{v},\quad   \cos\theta=\frac{\sqrt{2}v_H}{v}\ .
  \ee
and $v=174$ GeV. The mass of the triplet $A$ is given by 
   \be
   m_A^2=\frac{v_H^2+4 v_\Delta^2}{v_\Delta}\left(\lambda\left[2\mu-\mu_\Delta-(2\lambda-\lambda_3)v_\Delta\right]-A_\lambda\right)
   \label{masa1}
   \ee
 As we have seen from Fig.~\ref{ewbregion} one expects $v_\Delta< v_H$ and therefore it will be useful to provide the series expansion of the different masses and mixing angles in powers of $v_\Delta$. We will just present series expansions in powers of $v_\Delta$ to have an analytical feeling of the results although the numerical analysis will be done with the complete expressions. In particular for the squared mass in (\ref{masa1}) one can write the expansion
 \be
 m_A^2=\frac{v_H^2}{v_\Delta}\left[ \lambda(2\mu-\mu_\Delta)-A_\lambda\right]-\lambda(2\lambda-\lambda_3)v_H^2   +\mathcal O(v_\Delta)
 \ee

 There are also two scalar triplets $(T_H,T_\Delta)$ defined as
 \be
 T_H=\left(\begin{array}{c}
 \frac{1}{\sqrt{2}}(h_3^+ +h_3^{-*})\\
 h_{3R}^0\\
 \frac{1}{\sqrt{2}}(h_3^- +h_3^{+*})
 \end{array}\right),\quad T_\Delta=\left(\begin{array}{c}
 \frac{1}{\sqrt{2}}(\delta_3^+ +\delta_3^{-*})\\
 \delta_{3R}^0\\
\frac{1}{\sqrt{2}}( \delta_3^- +\delta_3^{+*})
 \end{array}\right)
\label{tripletes}
 \ee  
  which are mixed by the squared mass matrix $\mathcal M^2$  as
  \be 
  (T_H,T_\Delta)\mathcal M^2 \left(\begin{array}{c}T_H\\T_\Delta\end{array}\right)
 \label{masatripletes}
  \ee
  where
  \begin{align}
  \mathcal M^2_{11}&=  G^2v_H^2+2\lambda(-4\lambda v_\Delta^2+\lambda v_H^2+4\mu v_\Delta)+2 m_3^2-2v_\Delta[A_\lambda+\lambda(\mu_\Delta-\lambda_3 v_\Delta)]
  \nonumber\\
 \mathcal M^2_{22}&=\frac{4 G^2 v_\Delta^3-2\lambda v_H^2(\lambda v_\Delta-\mu)-v_\Delta[2 B_\Delta-3\lambda\lambda_3 v_H^2+2 v_\Delta(\lambda_3^2 v_\Delta-A_3)]-(\lambda v_H^2-2\lambda_3 v_\Delta^2)\mu_\Delta-v_H^2A_\lambda}{v_\Delta}
  \nonumber\\
  \mathcal M^2_{12}&=\mathcal M^2_{21}=2v_H\left[-A_\lambda+v_\Delta(G^2-4\lambda^2-\lambda\lambda_3)+\lambda\mu_\Delta\right]
  \label{masastripletes}
  \end{align}
In (\ref{masastripletes}) $G^2=g^2$ for the charged components and $G^2=g^2+g'^2$ for the neutral components. This just reflects the fact that the hypercharge coupling $g'$ breaks the custodial $SU(2)_V$ symmetry and thus spoils the triplet structure of $T_H$ and $T_\Delta$ and therefore of the two triplet mass eigenstates. The triplets eigenvectors are given by~\footnote{The presence of the $\mathcal O(g'^2)$ terms can be easily accounted for by just keeping in mind the different definition of $G^2$ for the mass eigenstates and mixing angles of the neutral and charged components of the triplets.}
 \be
\left(\begin{array}{c}T_1\\T_2\end{array} \right)= \left(\begin{array}{cc}\cos{\alpha_T} & -\sin{\alpha_T}\\ \sin{\alpha_T}&\cos{\alpha_T}\end{array}\right)\left(\begin{array}{c}T_H\\T_\Delta\end{array} \right)
 \ee
where the mixing angle $\alpha_T$ is given by
\be
\sin 2\alpha_T=\frac{2\mathcal M^2_{12}}{\sqrt{\tr ^2\mathcal M^2-4\det \mathcal M^2}},\quad \cos 2\alpha_T=\frac{\mathcal M^2_{22}-\mathcal M^2_{11}}{\sqrt{\tr^2 \mathcal M^2-4\det \mathcal M^2}}   
\label{mixingangles}
\ee   
and we are assuming that $m_{T_1}^2<m_{T_2}^2$. The expansion of the mass eigenvalues and the mixing angle in powers of $v_\Delta$ yields
\begin{align}
m_{T_1}^2&=G^2 v_H^2+2 m_3^2+2 \lambda^2 v_H^2+\mathcal O(v_\Delta)\nonumber\\
m_{T_2}^2&=\frac{v_H^2}{v_\Delta}\left[\lambda(2\mu-\mu_\Delta)-A_\lambda\right]-2B_\Delta-2\lambda^2v_H^2+3\lambda\lambda_3 v_H^2+\mathcal O(v_\Delta)\nonumber\\
\sin\alpha_T&=\frac{2(\lambda\mu_\Delta-A_\lambda)}{\lambda(2\mu-\mu_\Delta)-A_\lambda}\, \frac{v_\Delta}{v_H}+\mathcal O(v_\Delta^2)
\end{align}

\subsection{$SU(2)_V$ fiveplets}
The complex fiveplet in $\bar\Delta$ splits into two fiveplets:
a scalar fiveplet $F_{S}$ which contains the neutral scalar $\delta_{5R}^0$, and a pseudoscalar fiveplet $F_{P}$ which contains the neutral pseudoscalar $\delta_{5I}^0$. They are defined as
\be
F_S=\left(\begin{array}{c}\frac{1}{\sqrt{2}}(\delta_5^{++}+\delta_5^{--*})\\ \frac{1}{\sqrt{2}}(\delta_5^+ +\delta_5^{-*})\\ \delta_{5R}^0\\ \frac{1}{\sqrt{2}}(\delta_5^- +\delta_5^{+*})\\ \frac{1}{\sqrt{2}}(\delta_5^{--}+\delta_5^{++*})\end{array}
\right), \quad
F_P=\left(\begin{array}{c}\frac{1}{\sqrt{2}}(\delta_5^{--*}-\delta_5^{++})\\ \frac{1}{\sqrt{2}}(\delta_5^{-*}-\delta_5^+)\\ \delta_{5I}^0\\  \frac{1}{\sqrt{2}}(\delta_5^{+*}-\delta_5^-)\\ \frac{1}{\sqrt{2}}(\delta_5^{++*}-\delta_5^{--})\end{array}
\right) 
\ee
with masses squared
\begin{align}
m^2_{F_S}&=\frac{v_H^2(2\lambda\mu-6\lambda^2 v_\Delta^2-A_\lambda)+v_\Delta[3\lambda\lambda_3 v_H^2+2 v_\Delta(A_3-\lambda_3^2 v_\Delta)]-(\lambda v_H^2-6\lambda_3 v_\Delta^2)\mu_\Delta}{v_\Delta}
\nonumber\\
m^2_{F_P}&=\frac{v_H^2(2\lambda\mu-6\lambda^2 v_\Delta^2+\lambda\lambda_3 v_\Delta-A_\lambda)-2 v_\Delta B_\Delta-(\lambda v_H^2-4\lambda_3 v_\Delta^2)\mu_\Delta}{v_\Delta}\ .
\end{align}
The power expansion in $v_\Delta$ of $m^2_{F_S}$ and $m^2_{F_P}$ reads as
\begin{align}
m^2_{F_S}&=\frac{v_H^2}{v_\Delta}\left[ \lambda(2\mu-\mu_\Delta)-A_\lambda \right]-3\lambda(2\lambda-\lambda_3)v_H^2+\mathcal O(v_\Delta)
\nonumber\\
m^2_{F_P}&=\frac{v_H^2}{v_\Delta}\left[ \lambda(2\mu-\mu_\Delta)-A_\lambda \right]-2 B_\Delta-\lambda(6\lambda-\lambda_3)v_H^2+\mathcal O(v_\Delta)
\end{align}

\subsection{$SU(2)_V$ singlets}
Moreover there are in the spectrum two real neutral scalar $(h_{1R}^0,\delta_{1R}^0)$ singlets mixed by the mass matrix $\mathcal M^2_S$
  \be 
  (h_{1R}^0,\delta_{1R}^0)\mathcal M^2_S \left(\begin{array}{c}h_{1R}^0\\ \delta_{1R}^0\end{array}\right)
 \label{masasingS}
  \ee
where
\begin{align}
(\mathcal M^2_S)_{11}&=6\lambda^2 v_H^2
\nonumber\\
(\mathcal M^2_S)_{22}&=\frac{v_H^2\left[\lambda(2\mu-\mu_\Delta)-A_\lambda\right]+v_\Delta^2\left[-A_3+\lambda_3(4\lambda_3 v_\Delta-3\mu_\Delta)  \right]}{v_\Delta}
\nonumber\\
(\mathcal M^2_S)_{12}&=(\mathcal M^2_S)_{21}=\sqrt{6}v_H\left[A_\lambda+\lambda(6\lambda v_\Delta-2\lambda_3 v_\Delta-2\mu+\mu_\Delta
)\right] \ .
\end{align}
The eigenvectors can be written in terms of the rotation with angle $\alpha_{S}$ as
 \be
\left(\begin{array}{c}S_1\\S_2\end{array} \right)= \left(\begin{array}{cc}\cos{\alpha_S} & -\sin{\alpha_S}\\\sin{\alpha_S}&\cos{\alpha_S}\end{array}\right)\left(\begin{array}{c}h_{1R}^0\\ \delta_{1R}^0\end{array} \right)
 \ee
where the mixing angle $\alpha_S$ is defined as in Eq.~(\ref{mixingangles}) and we are assuming that $m_{S_1}^2<m_{S_2}^2$. The expansion of the mass eigenvalues and the mixing angle in powers of $v_\Delta$ yields
\begin{align}
m_{S_1}^2&=6\lambda^2 v_H^2+\mathcal O(v_\Delta)
\nonumber\\
m_{S_2}^2&=\frac{\lambda(2\mu-\mu_\Delta)-A_\lambda}{v_\Delta}+\mathcal O(v_\Delta)
\nonumber\\
\sin\alpha_S&=-\sqrt{6}\frac{v_\Delta}{v_H}+\mathcal O(v_\Delta^2)
\label{S}
\end{align}
From Eq.~(\ref{S}) we can see that the scalar singlet $S_1$ plays the role (in the limit where $v_\Delta\to 0$) of the light $CP$-even MSSM Higgs $h$ when decoupled triplets are added in the superpotential~\cite{Delgado:2012sm}.

Finally there are also two pseudoscalar singlets $(h_{1I}^0,\delta_{1I}^0)$ mixed by the mass matrix $\mathcal M^2_P$
  \be 
  (h_{1I}^0,\delta_{1I}^0)\mathcal M^2_P \left(\begin{array}{c}h_{1I}^0\\ \delta_{1I}^0\end{array}\right)
 \label{masasingP}
  \ee
where

\begin{align}
(\mathcal M^2_P)_{11}&=2\left(m_3^2-3 v_\Delta\left[ A_\lambda+\lambda(-\lambda_3 v_\Delta+\mu_\Delta) \right]   \right)
\nonumber\\
(\mathcal M^2_P)_{22}&=-\frac{v_H^2(A_\lambda-2\lambda\mu)+v_\Delta(-3A_3v_\Delta+2 B_\Delta-4\lambda\lambda_3 v_H^2)+(\lambda v_H^2-\lambda_3 v_\Delta^2)\mu_\Delta}{v_\Delta}
\nonumber\\
(\mathcal M^2_P)_{12}&=(\mathcal M^2_P)_{21}=\sqrt{6}v_H\left[ \lambda(-2\lambda_3 v_\Delta+\mu_\Delta)-A_\lambda \right]
\end{align}
The eigenvectors can be written in term of the rotation with angle $\alpha_{P}$ as
 \be
\left(\begin{array}{c}P_1\\P_2\end{array} \right)= \left(\begin{array}{cc}\cos{\alpha_P} & -\sin{\alpha_P}\\ \sin{\alpha_P}&\cos{\alpha_P}\end{array}\right)\left(\begin{array}{c}h_{1I}^0\\ \delta_{1I}^0\end{array} \right)
 \ee
where again the mixing angle $\alpha_P$ is defined as in Eq.~(\ref{mixingangles})  and we are assuming that $m_{P_1}^2<m_{P_2}^2$. The expansion of the mass eigenvalues and the mixing angle in powers of $v_\Delta$ yields
\begin{align}
m_{P_1}^2&=2m_3^2+\mathcal O(v_\Delta)
\nonumber\\
m_{P_2}^2&=\frac{v_H^2\left[\lambda(2\mu-\mu_\Delta)-A_\lambda \right]}{v_\Delta}-2B_\Delta+4\lambda\lambda_3 v_H^2+\mathcal O(v_\Delta)
\nonumber\\
\sin\alpha_P&=\frac{2(\lambda\mu_\Delta-A_\lambda)}{\lambda(2\mu-\mu_\Delta)-A_\lambda}\, \frac{v_\Delta}{v_H}+\mathcal O(v_\Delta^2)
\label{P}
\end{align}
Notice also from Eq.~(\ref{P}) that the pseudoscalar $P_1$ plays the role (in the limit $v_\Delta\to 0$) of the massive MSSM pseudoscalar.

 \subsection{Numerical analysis}
 
 In this section we will provide plots of the mass eigenvalues previously obtained in this section. We will make the choice of parameters used in the left panel of Fig.~\ref{ewbregion}, i.e.
\be
A_\lambda=A_{\lambda_3}=0,\  \mu=\mu_\Delta=250\ \textrm{GeV},\  m_3=500\ \textrm{GeV},\  B_\Delta=-m_3^2,\  \lambda_3=-0.35\ .
\label{valoresfinales}
\ee
As for the value of $\lambda$ we will trade it for the mass $m_{S_1}\simeq 126$ GeV by identifying the mass eigenstate $S_1$ with the recently discovered Higgs-like particle at the LHC. As the relevant values of $\lambda$ are moderately small we will consistently neglect the radiative corrections to the mass eigenvalue arising from the $\lambda$ coupling and those from the bottom quark Yukawa coupling (as $\tan\beta=1$), and will only keep those coming from the top Yukawa coupling. Moreover as we are neglecting trilinear soft supersymmetry breaking terms in (\ref{valoresfinales}) we will do so for the trilinear coupling $A_t$ in the stop sector which we will neglect. On the other hand this choice is the most conservative one as, in the absence of threshold corrections from the stop sector, the only tree level contribution to the Higgs mass comes from the coupling $\lambda$ which (along with the leading radiative corrections) has to cope with the experimental value of the Higgs mass. Therefore by including the leading $(\propto h_t^2)$ one-loop corrections and the subleading two-loop QCD corrections $(\propto h_t^2\alpha_3)$~\cite{Carena:1995bx} one obtains for  $m_{\widetilde t}\simeq 650\,$GeV that radiative corrections amount to a contribution $\simeq (72\,\textrm{GeV})^2$ to the squared Higgs mass which leaves a tree-level squared mass contribution $\simeq (104\,\textrm{GeV})^2$.
\begin{figure}[htb]
\begin{center}
\vspace{3mm}
\includegraphics[width=80mm]{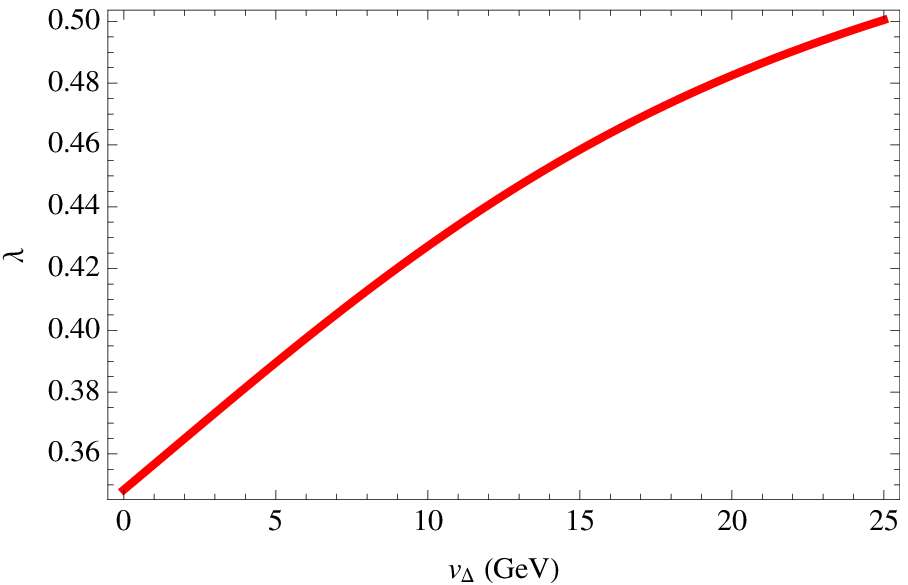}
\includegraphics[width=82mm]{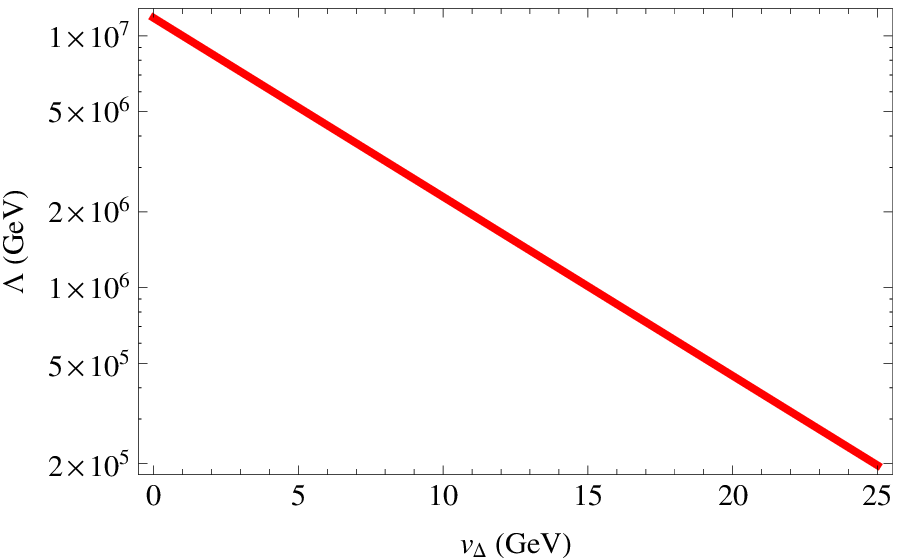}
\end{center}
\caption{\it Left panel: Plot of $\lambda$ as a function of $v_\Delta$ for the values of supersymmetric parameters in Eq.~(\ref{valores}), $\lambda_3=-0.35$ and $m_{S_1}\simeq 126$ GeV. Right panel:  Plot of the scale at which the couplings reach Landau poles as a function of $v_\Delta$ for the (initial) values of $\lambda(m_t)$ as given in the left panel.}
\label{lambda}
\end{figure}
The corresponding value of $\lambda$ is plotted in Fig.~\ref{lambda} as a function of $v_\Delta$. By comparison with the left plot of Fig.~\ref{ewbregion} we can see that the parameter space selected by the value of the Higgs mass is entirely consistent with electroweak breaking.

A very general feature of this model is that it requires an ultraviolet (UV) completion as the dimensionless couplings reach Landau poles for scales below the Planck scale as an effect of the renormalization group running. In turn radiative corrections provided by the top quark Yukawa coupling and by $g'$ break custodial invariance of dimensionless couplings at high scales. A detailed discussion and technical details of this issue can be found in appendix~\ref{RGE}. In the right panel of Fig.~\ref{lambda} we plot the value of the Landau pole for the couplings as a function of $v_\Delta$ for the custodial values of the parameters at $m_t$ given in Eq.~(\ref{valores}). The origin of the existence of the Landau pole is three-fold:
\begin{itemize}
\item
The very existence of three $SU(2)_L$ triplets makes the weak coupling $g$ become non-perturbative at one-loop at a scale $\sim 10^{13}$ GeV.
\item
The larger $v_\Delta$, the larger the top Yukawa coupling $h_t(m_t)$, as it has to compensate for the smallness of the Higgs VEV $v_H$ which couples to the top quark. This by itself should put a bound on the value of $v_\Delta$ as the raw bound $h_t(m_t)\lesssim 4\pi$ already translates into the bound $v_\Delta\lesssim 86$ GeV.
\item
The starting value of $\lambda(m_t)$ has to cope with the experimental value of the Higgs mass $m_h=126$ GeV after considering the top-stop sector radiative corrections to the Higgs mass. We have conservatively assumed zero mixing in the stop sector $A_t\simeq 0$ so that for other values of the mixing (as e.g.~for maximal mixing $A_t\simeq \sqrt{6} m_Q$) the initial value $\lambda(m_t)$ can be decreased and thus the location of the Landau pole moved away.
\end{itemize}
As a consequence of this behavior we conclude that the considered model requires low-scale supersymmetry breaking as we will also discuss in section~\ref{breaking}.

\begin{figure}[htb]
\begin{center}
\vspace{3mm}
\includegraphics[width=80mm]{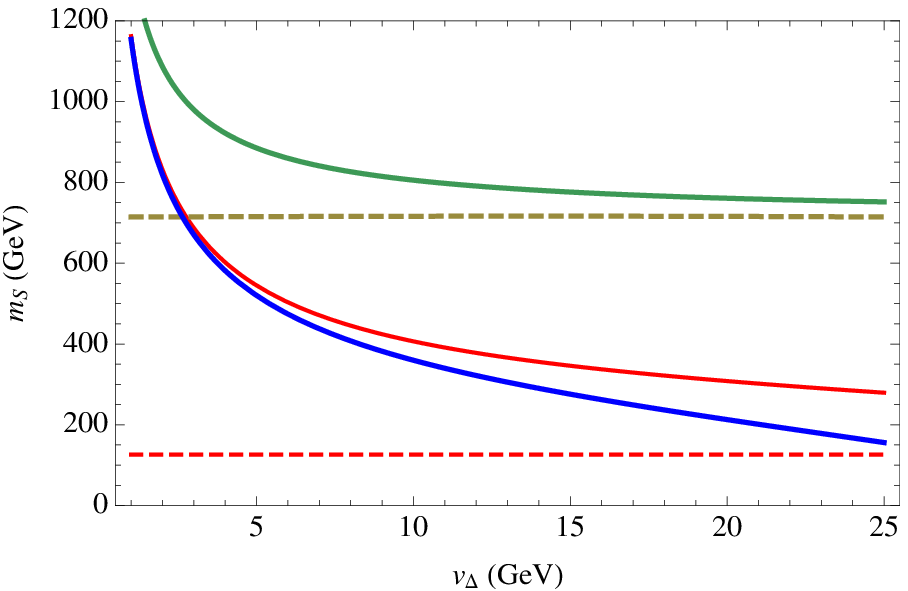}
\includegraphics[width=80mm]{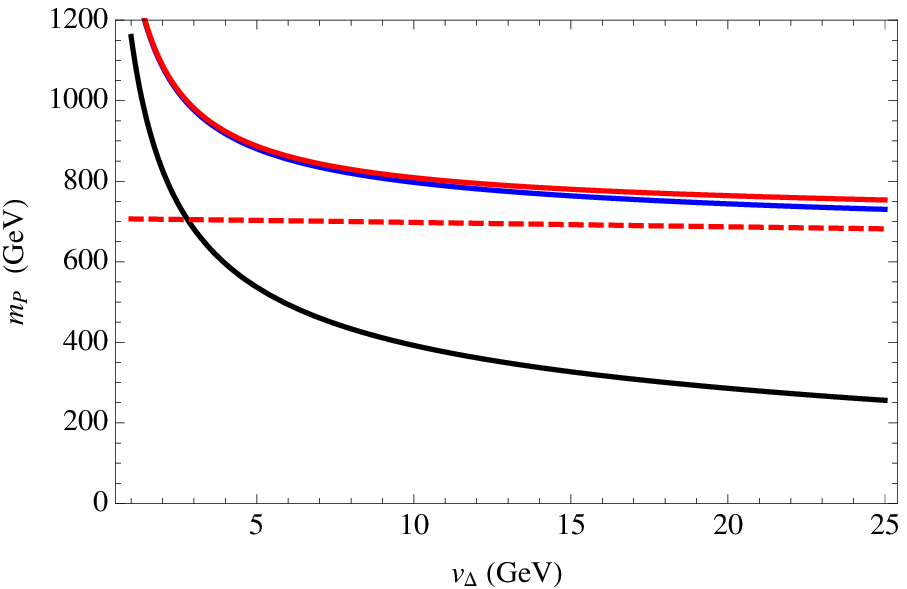}
\end{center}
\caption{\it Left panel: Masses (in GeV) of scalar multiplets as a function of $v_\Delta$. From bottom-up the different lines correspond to the mass eigenstates: $S_1,\, F_S,\, S_2,\, T_1,\, T_2$. Right panel: The same for the pseudoscalar multiplets. From bottom-up the different lines correspond to the mass eigenstates: $A,\, P_1,\, F_P,\, P_2$. 
}
\label{masas}
\end{figure}

The masses of the eigenstates studied earlier on in this section are provided in Fig.~\ref{masas}
 for the choice of supersymmetric parameters given in Eq.~(\ref{valoresfinales}). We are using the same color code for the mass eigenstates which are mixed, through a mixing angle, from the original interaction states: in all cases the eigenstates which decouple in the $v_\Delta\to 0$ limit are presented in solid lines and their companions in dashed lines. 
 As we can see from Fig.~\ref{masas} $S_1$ (the SM-like Higgs) is the lightest scalar and the second to lightest scalar is $F_S$ which is supermassive for $v_\Delta\to 0$ but becomes as massive as $S_1$ for $v_\Delta\simeq 25$ GeV. We could think that this is in conflict with present experimental data for the present choice of supersymmetric parameters. However, as we will see in the following section, this state will be weakly coupled to gauge bosons, as its couplings are suppressed by $\sin\theta$, and it is not coupled at all to fermions. The third to lightest scalar is $S_2$ which is supermassive in the limit $v_\Delta\to 0$ and whose mass becomes $\simeq 300$ GeV for $v_\Delta\simeq 25$ GeV. Unlike $F_S$ this state is coupled to both gauge bosons and fermions but, as it is the orthogonal combination to the SM-like Higgs $S_1$, it is weakly coupled and should not be easily detected at LHC. Finally the heaviest scalars, $T_1$ and $T_2$ are superheavy. Similarly the lightest pseudoscalar is $A$, the orthogonal combination to the Goldstone bosons, whose mass grows when $v_\Delta\to 0$. In the region $v_\Delta\sim 25$ GeV its mass is $m_A\simeq 250$ GeV. It does not couple to gauge bosons and its coupling to fermions is suppressed by $\sin\theta$. Notice that rates with couplings proportional to $\sin\theta$ are suppressed by $\sin^2\theta\lesssim 0.16$ for $v_\Delta\lesssim 25$ GeV.
 
 \subsection{General considerations on decoupling regions}
From the previous results in this section it is clear that there are two decoupling limits in the Higgs sector:
\begin{itemize}
\item
The limit $m_\Delta\to \infty$ (i.e.~$v_\Delta\to 0$) in which case all states arising from the triplet $\bar\Delta$ are very heavy and decouple from the Higgs sector in the doublet $\bar H$. In this limit the heavy states are the scalar $S_2$ and pseudoscalar $P_2$ singlets, the scalar $T_2$ and pseudoscalar $A$ triplets and the scalar $F_S$ and pseudoscalar $F_P$ fiveplets. Similarly the light states are (apart from the massless Goldstone triplet $G$) the scalar $S_1$ and pseudoscalar $P_1$ singlets and the scalar triplet $T_1$. As $v_\Delta< v_H$ we expect this limit to provide an approximate classification of the states.
\item
The limit $m_H\to\infty$ (i.e. $m_3^2\to\infty$) in which case the previous light states split into heavy and light states. Heavy states, with masses controlled by the supersymmetry breaking parameter $m^2_H$, i.e.~by $m_3^2$, are the pseudoscalar singlet $P_1$ and the scalar triplet $T_1$. In particular $P_1$ plays the role of the massive MSSM pseudoscalar, and the scalar triplet $T_1$ that of massive neutral and charged Higgses in the MSSM with decoupled triplets. Finally the only light scalar (not controlled by the supersymmetry breaking scale) is the scalar singlet $S_1$ which plays the role of the MSSM light SM-like Higgs.
 \end{itemize}

 \section{The fermion sector}
 \label{fermionsector}
 In this section we will present the mass matrices for fermions in the Higgs doublet-triplet mixed sectors.

\subsection{Neutralinos}
In the basis $(\tilde B,\tilde W_3,\tilde H_1^0,\tilde H_2^0,\tilde\phi^0,\tilde \chi^0,\tilde \psi^0)$ the neutralino Majorana mass matrix is given by
\be
\mathcal M_{1/2}^0=
\left(\begin{array}{ccccccc}
M_1 & 0& -\frac{g'v_H}{\sqrt{2}}&\frac{g'v_H}{\sqrt{2}}& 0 &-\sqrt{2} g'v_\Delta &\sqrt{2} g' v_\Delta\\
0&M_2 &\frac{gv_H}{\sqrt{2}} &-\frac{gv_H}{\sqrt{2}}  &0 &\sqrt{2}g v_\Delta &-\sqrt{2}g v_\Delta \\
-\frac{g'v_H}{\sqrt{2}} &\frac{gv_H}{\sqrt{2}}  &2\lambda v_\Delta &\lambda v_\Delta-\mu &\lambda v_H & 0&2\lambda v_H \\
\frac{g'v_H}{\sqrt{2}} &-\frac{gv_H}{\sqrt{2}}  &\lambda v_\Delta-\mu &2\lambda v_\Delta &\lambda v_H &2\lambda v_H &0 \\
0& 0&\lambda v_H &\lambda v_H &\mu_\Delta &-\lambda_3 v_\Delta &-\lambda_3 v_\Delta \\
-\sqrt{2} g' v_\Delta&\sqrt{2} g v_\Delta &0 &2\lambda v_H &-\lambda_3 v_\Delta &0 &-\lambda_3 v_\Delta+\mu_\Delta \\
\sqrt{2} g' v_\Delta&-\sqrt{2}g v_\Delta &2\lambda v_H &0 &-\lambda_3 v_\Delta&-\lambda_3 v_\Delta +\mu_\Delta& 0\\
\end{array}
\right)
\ee
\begin{figure}[htb]
\begin{center}
\includegraphics[width=80mm]{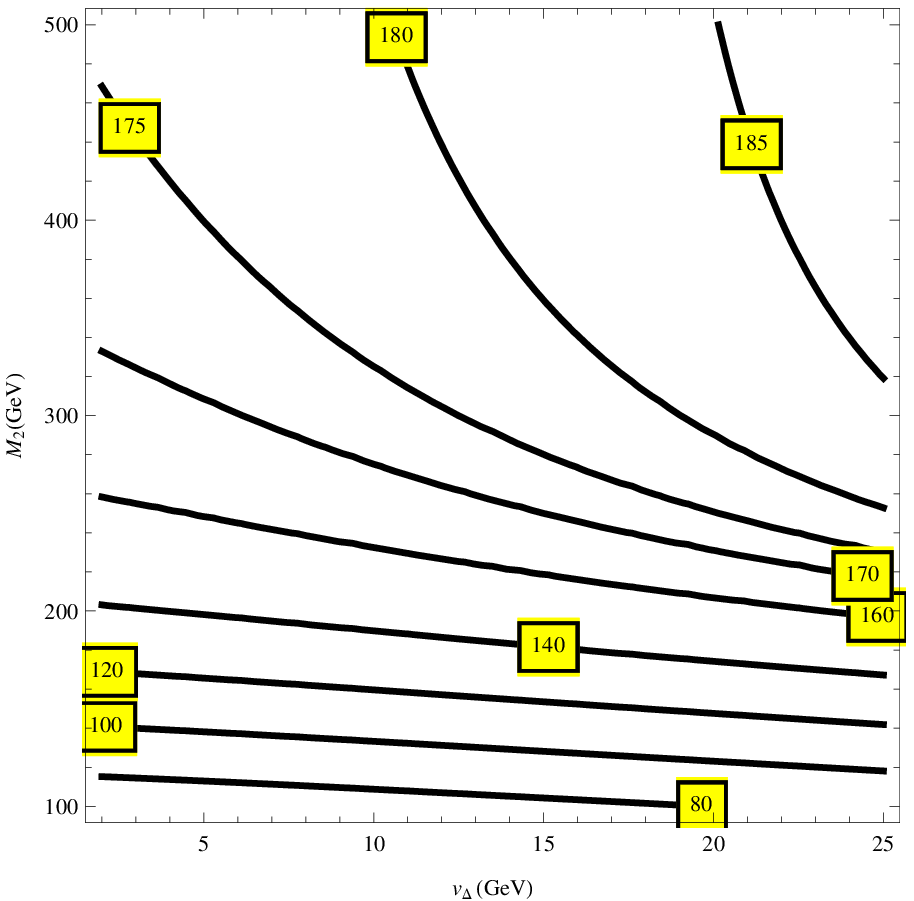}
\includegraphics[width=80mm]{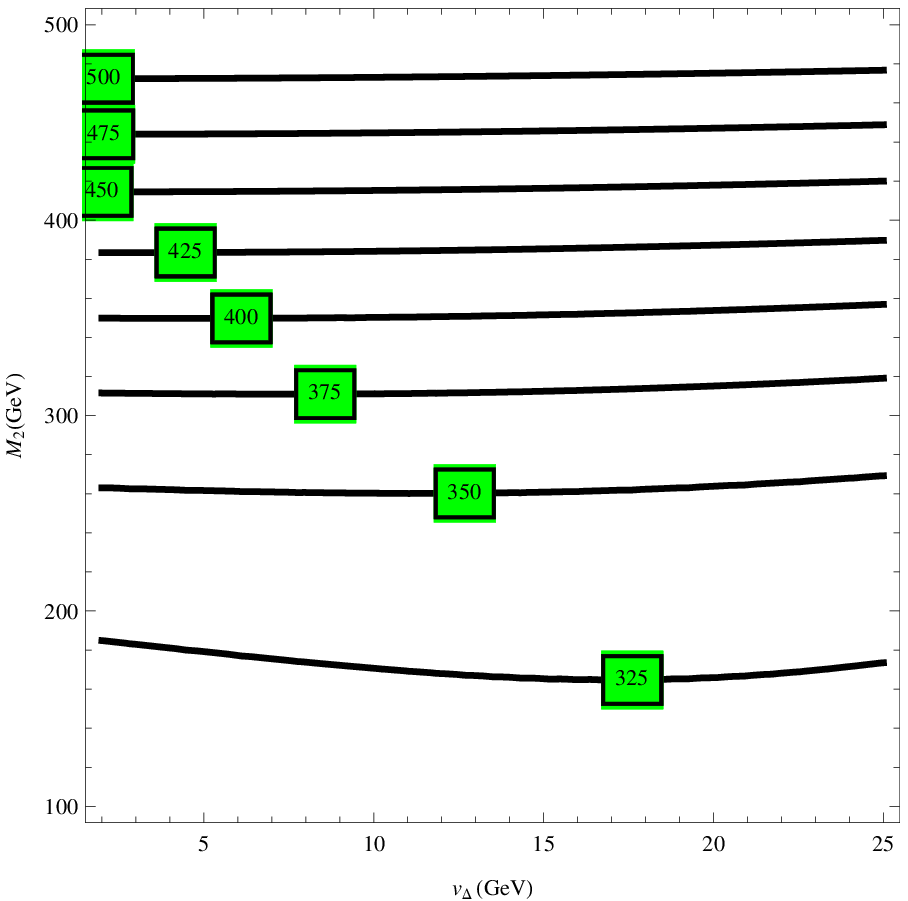}
\end{center}
\caption{\it Contour plots in the plane $(v_\Delta,M_2)$ of the lightest (left panel) and heaviest (right panel) neutralino masses for the values of the supersymmetric parameters in Eq.~(\ref{valores}) and $M_1=200$ GeV. All masses are in GeV}
\label{neutralinos}
\end{figure}
In Fig.~\ref{neutralinos} we make contour plots in the plane $(v_\Delta,M_2)$ of the lightest neutralino mass eigenvalue (left panel) and the heaviest neutralino mass eigenvalue (right panel), both in GeV, for the values of parameters in Eq.~(\ref{valores}) and $M_1=200$ GeV.

 \subsection{Charginos}
\begin{figure}[htb]
\begin{center}
\includegraphics[width=80mm]{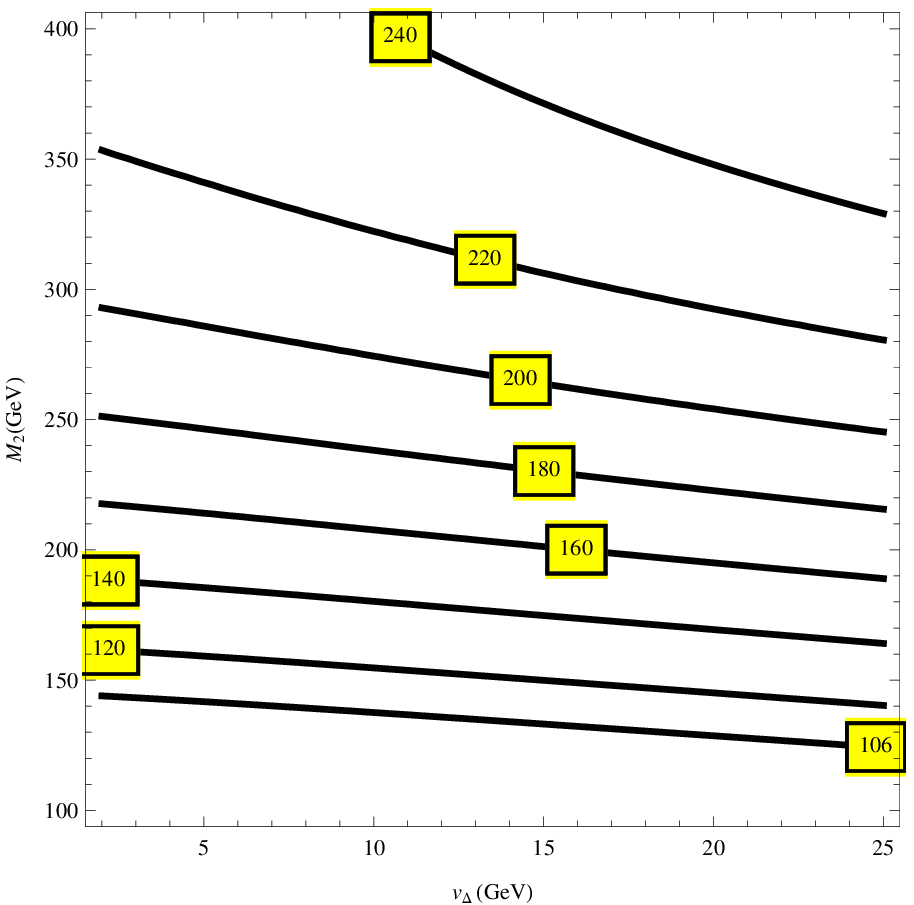}
\includegraphics[width=80mm]{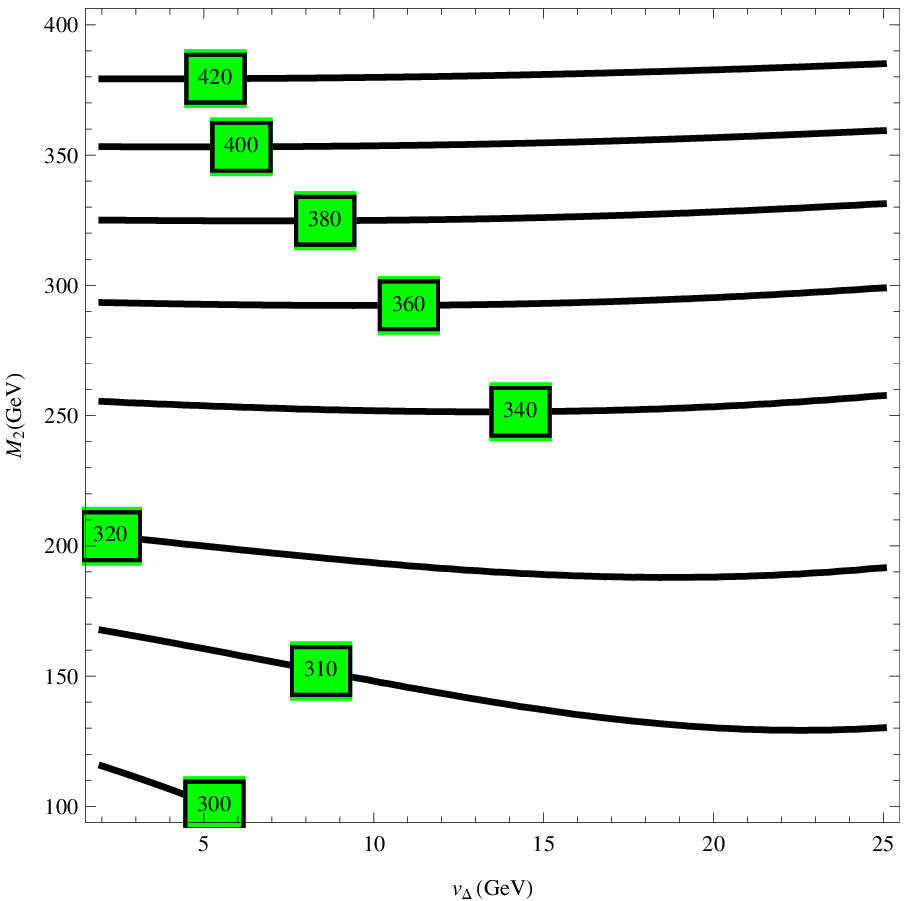}
\end{center}
\caption{\it Contour plots in the plane $(v_\Delta,M_2)$ of the lightest (left panel) and heaviest (right panel) chargino masses for the values of the supersymmetric parameters in Eq.~(\ref{valores}). All masses are in GeV.}
\label{charginos}
\end{figure}

 The mass Lagrangian for charginos with $Q_f=\pm 1$  in the basis $\Psi^+=(\tilde W^+,\tilde H_2^+,\tilde \phi^+,\tilde \psi^+)^T$ and $\Psi^-=(\tilde W^-,\tilde H_1^-,\tilde \phi^-,\tilde \chi^-)^T$ is given by
\be
\mathcal L_{1/2}^\pm=-\frac{1}{2} 
\left(\begin{array}{cc} \Psi^{+T} & \Psi^{-T}\end{array}
\right)
\left(\begin{array}{cc} 0 & \mathcal M_{1/2}^{\pm T}\\ \mathcal M_{1/2}^\pm & 0 \end{array}
\right)
\left(\begin{array}{c} \Psi^+\\\Psi^-\end{array}
\right)+ h.c.
\ee
 where
 \be
 \mathcal M_{1/2}^{\pm}=
 \left(
 \begin{array}{cccc}
 M_2&g v_H &\sqrt{2} g v_\Delta &-\sqrt{2} g v_\Delta \\
g v_H &\lambda v_\Delta+\mu &-\sqrt{2}\lambda v_H &\sqrt{2}\lambda v_H \\
-\sqrt{2} g v_\Delta &\sqrt{2} \lambda v_H &\mu_\Delta &\lambda_3 v_\Delta \\
\sqrt{2} g v_\Delta & -\sqrt{2} \lambda v_\Delta&\lambda_3 v_\Delta& \mu_\Delta 
 \end{array}
 \right)
 \ee
 In Fig.~\ref{charginos} we make contour plots in the plane $(v_\Delta,M_2)$ of the simply charged lightest chargino mass eigenvalue (left panel) and the heaviest chargino mass eigenvalue (right panel) for the values of parameters in Eq.~(\ref{valores}).

 Finally for the doubly charged fermions $(\tilde\psi^{++}, \tilde\chi^{--})$ the Dirac mass is given by
 \be
 \mathcal M_{1/2}^{\pm\pm}=\mu_\Delta-\lambda_3 v_\Delta
 \ee
 which using the chosen set of parameters in Eq.~(\ref{valores}) is $\mathcal M_{1/2}^{\pm\pm}\simeq \mu_\Delta=250$ GeV.

\section{Unitarity}
\label{unitarity}
 Perturbative unitarity translates into bounds on scattering amplitudes involving longitudinally polarized gauge bosons~\cite{Lee:1977eg}. In particular the condition to achieve perturbative unitarity is that tree-level scattering amplitudes $V_LV_L\to V_LV_L$, where $V_L$ denotes the longitudinal polarization of the gauge boson $V$, mediated by the Higgs sector reproduce at high energy $(s\to\infty)$ the SM behavior. These amplitudes  involve the trilinear couplings $g_{\mathcal H VV}$ where $\mathcal H$ goes over the set of mass eigenstates described in section~\ref{Higgssector}. The relevant couplings are listed here~\footnote{We are missing intentionally a global factor $i\eta_{\mu\nu}$.}:
 \begin{align}
 g_{F_S^{++}W^-W^-}&=g_{F_S^{--}W^+W^+}=-\sqrt{2}gm_W \sin \theta \nonumber\\
g_{S_1W^+W^-}&= gm_W\left( \cos\theta \cos\alpha _S-\sqrt{\frac{8}{3}} \sin\theta \sin\alpha _S\right)\nonumber\\
g_{S_2W^+W^-}&= gm_W\left( \cos\theta \sin\alpha _S+\sqrt{\frac{8}{3}} \sin\theta \cos\alpha _S\right)\nonumber\\
g_{F_S^0 W^+W^-}&=-\frac{gm_W \sin\theta}{\sqrt{3}}\nonumber\\
g_{S_1ZZ}&=\frac{gm_W}{\cos^2\theta_W} \left(\cos\theta \cos\alpha _S-\sqrt{\frac{8}{3}} \sin \theta \sin\alpha _S\right)\nonumber\\
g_{S_2ZZ}&=\frac{gm_W}{\cos^2\theta_W} \left(\cos\theta \sin\alpha _S+\sqrt{\frac{8}{3}}  \sin \theta \cos\alpha _S\right)\nonumber\\
g_{F_S^0ZZ}&=\frac{2gm_W  \sin \theta }{\sqrt{3}\cos^2\theta_W}\nonumber\\
g_{F_S^-W^+Z}&=g_{F_S^+W^-Z}=-\frac{gm_W \sin\theta}{\cos\theta_W}\nonumber\\
g_{G^-W^+Z}&=-g_{G^+W^-Z}=-gm_W \sin\theta_W \tan\theta_W
\label{trilineales}
 \end{align}

 In order to exhibit how unitarity works in this model we will choose two particular amplitudes: the elastic scatterings $Z_LW_L^+\to Z_L W_L^+$ and $W_L^+ W_L^+\to W_L^+ W_L^+$.
 
 \subsection{The scattering $Z_LW_L^+\to Z_L W_L^+$}
 
 In this reaction the SM Higgs $h$ contributes in the $t$-channel and therefore in the limit $s\to\infty$ the amplitude is proportional to $t$ with the coupling
 \be
 g^{SM}_{hWW}\cdot g^{SM}_{hZZ} =g^2 m_W^2/\cos^2\theta_W 
\label{SM1coup}
 \ee

 In the supersymmetric custodial triplets model (SCTM) there are neutral scalars $\mathcal H^0_i=S_1,S_2,F_S^0$ which contribute in the $t$-channel and provide an amplitude proportional to $t$ in the limit where $s\to\infty$. On the other hand $F_S^+$ is exchanged in the $s$ and $u$-channels and provides amplitudes proportional to $s+u\simeq -t$. Therefore the total amplitude is proportional to $t$ with a coupling equal to
 \be
 \sum_{\mathcal H^0_i=S_1,S_2,F_S^0} g_{\mathcal H_i^0 W^+W^-}\cdot g_{\mathcal H_i^0 ZZ}-g_{F_S^+ W^-Z}^2
 \label{SCT1coup}
 \ee
 Now we can see that using Eqs.~(\ref{trilineales}) the coupling in Eq.~(\ref{SCT1coup}) reproduces the SM coupling of Eq.~(\ref{SM1coup}). 
 
 \subsection{The scattering $W_L^+ W_L^+\to W_L^+ W_L^+$}
 
 In this reaction the SM Higgs $h$ contributes in the $t$ and $u$ channels so that in the limit where $s\to\infty$ the amplitude is proportional to $t+u$ with a coupling
 \be
 (g^{SM}_{hWW})^2 =g^2 m_W^2
\label{SM2coup}
 \ee

 Similarly to the previous amplitude, in the SCTM the neutral scalars $\mathcal H_i^0$ contribute to the $t$ and $u$ channels with an amplitude, in the limit $s\to\infty$, proportional to $t+u\simeq -s$. Moreover the doubly charged scalar $F_S^{++}$ is exchanged in the $s$ channel with an amplitude proportional to $s$. Adding up the four terms one gets an amplitude which, in the asymptotic limit, is proportional to the coupling
 \be
 \sum_{\mathcal H^0_i=S_1,S_2,F_S^0} g_{\mathcal H_i^0 W^+W^-}^2-g_{F_S^{++} W^-W^-}^2
 \label{SCT2coup}
 \ee
 which, using the actual values of the couplings in (\ref{trilineales}), reproduces the SM result, Eq.~(\ref{SM2coup}).

 \section{The neutral Higgs rates}
 \label{Higgscouplings}
In this section we will study the CP-even neutral Higgs ($\mathcal H=S_1,\,S_2,\,F_S^0,\, T_1,\, T_2$) rates to a pair of gauge bosons $VV$ $(V=W,Z,\gamma)$ and SM fermions $ff$ $(f=t,b,\tau)$. We will first consider tree-level processes.
 
 \subsection{Tree-level rates}
  \label{Higgscouplings}
 The angles $\alpha_S$ and $\theta$ play a fundamental role in the interactions of the
CP-even Higgses contained in $SU(2)_V$ singlets $(S_1,\, S_2)$ with the SM fields, as shown in Eq.~(\ref{trilineales}) and Table~\ref{tabla1}.  
 \begin{table}[h!]
\begin{center}
\begin{tabular}{||c|c|c|c||}
\hline
\rule{0pt}{4mm}
$r_{S_1VV}$&$r_{S_1ff}$&$r_{S_2VV}$&$r_{S_2ff}$\\[1mm]
\hline\hline
\rule{0pt}{3.5ex}    
$\cos\alpha_S\cos\theta-\sqrt{\frac{8}{3}}\sin\alpha_S\sin\theta$&${\displaystyle \frac{\cos\alpha_S}{\cos\theta}}$&$\sin\alpha_S\cos\theta+\sqrt{\frac{8}{3}}\cos\alpha_S\sin\theta$&${\displaystyle \frac{\sin\alpha_S}{\cos\theta}}$\\[2mm]
\hline    
\end{tabular}
\end{center}
\caption{\it Ratios (\ref{ratios}) for the different channels.}
\label{tabla1}
\end{table}

\noindent The ratios $r_{\mathcal H XX}$ are the quantities
\be
r_{\mathcal H XX}=\frac{g_{\mathcal H XX}}{g_{h XX}^{\rm SM}}\quad \mathcal H=S_1,S_2\quad
{\rm with}\quad
X=V(W,Z),f(t,b,\tau)
\label{ratios}
\ee
\begin{table}[htb]
\begin{center}
\begin{tabular}{||c|c|c|c|c|c|c|c||}
\hline
\rule{0pt}{4mm}
$r_{T_i^0VV}$&$r_{T_1^0uu}$&$r_{T_1^0dd}$&$r_{T_2^0uu}$&$r_{T_2^0dd}$ &$r_{F_S^0WW}$ &$r_{F_S^0 ZZ}$&$ r_{F_S^0ff}$ \\[1mm]
\hline\hline
\rule{0pt}{3.5ex}    
$0$&${\displaystyle -\frac{\cos\alpha_T}{\cos\theta}}$&${\displaystyle \frac{\cos\alpha_T}{\cos\theta}}$&${\displaystyle -\frac{\sin\alpha_T}{\cos\theta}}$&${\displaystyle \frac{\sin\alpha_T}{\cos\theta}}$& ${\displaystyle  \frac{\sin\theta}{\sqrt{3}} }$ &${\displaystyle - \frac{2\sin\theta}{\sqrt{3}} }$& 0\\[2mm]
\hline    
\end{tabular}
\end{center}
\caption{\it Ratios for $T_i^0$ and $F_S^0$ in the different channels. Notation is $u=t$, $d=b,\tau$.}
\label{tabla2}
\end{table}
where $g_{\mathcal H XX}$ and $g_{h XX}^{SM}$ are the couplings between the Higgs $\mathcal H$ and the field $X$ in the SCTM and in the SM, respectively.
Similarly the couplings of the $CP$-even scalars contained in the triplets $T_i\ (i=1,2)$ and fiveplets $F_S^0$ are controlled by the angles $\alpha_T$ and $\theta$ as given in Table~\ref{tabla2}.

\begin{figure}[h!]
\begin{center}
\includegraphics[width=80mm]{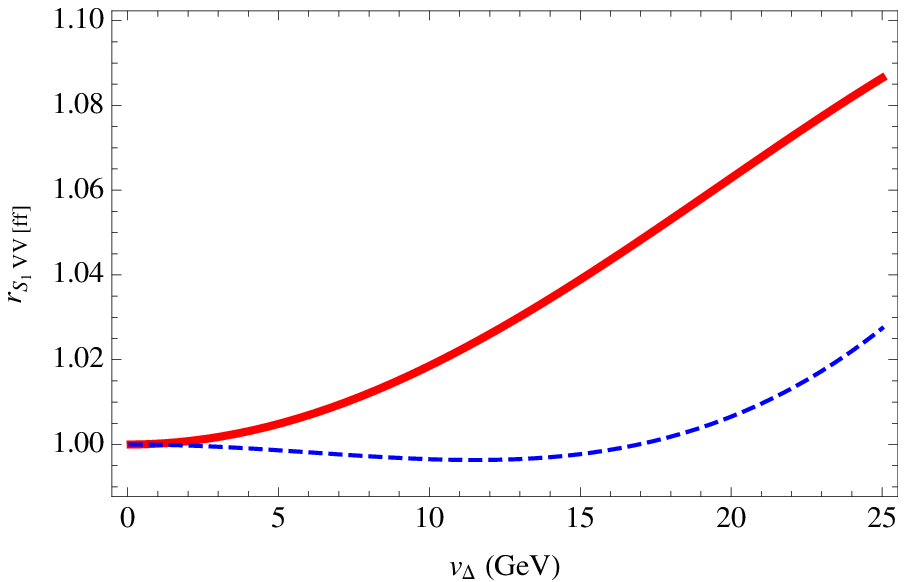}
\includegraphics[width=80mm]{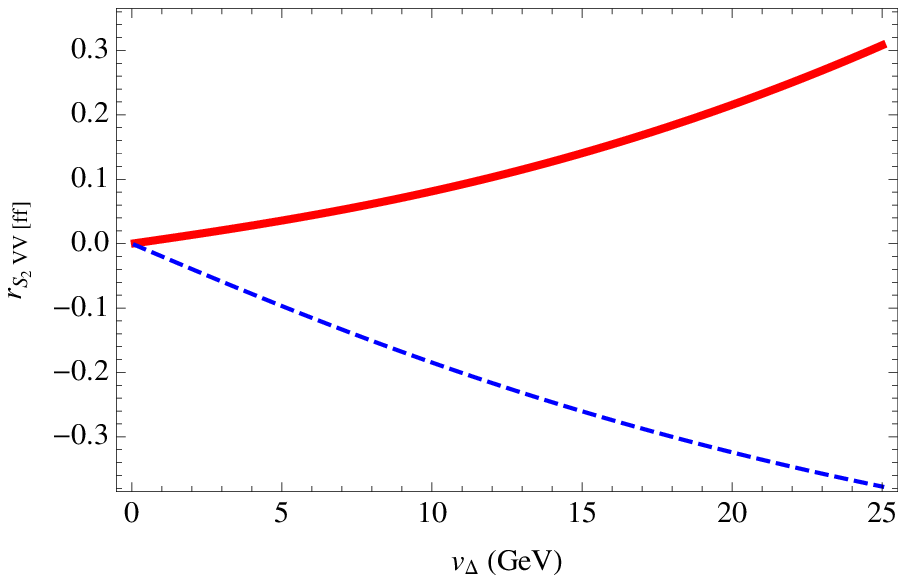}
\end{center}
\caption{\it Left panel: Plots of the ratios $r_{S_1VV}$ (upper solid line) and $r_{S_1 ff}$ (lower dashed line) as functions of $v_\Delta$. Right panel: Plots of the ratios $r_{S_2VV}$ (upper solid line) and $r_{S_2 ff}$ (lower dashed line) as functions of $v_\Delta$.}
\label{couplings}
\end{figure}

We plot in Fig.~\ref{couplings} the ratios of the SM-like Higgs $S_1$ couplings to gauge bosons $r_{S_1 VV}$ (left panel, solid line) and fermions $r_{S_1 ff}$ (left panel, dashed line). We also present the corresponding couplings of its orthogonal partner $S_2$ couplings 
to gauge bosons $r_{S_2 VV}$ (right panel, solid line) and fermions $r_{S_2 ff}$ (right panel, dashed line). As we can see from the left panel of Fig.~\ref{couplings} the couplings of $S_1$ to $VV$ and $ff$ are in very good agreement with the 68\% CL intervals on $r_{hVV}$ and $r_{hff}$ (when profiling over the other parameter) as measured for instance by the ATLAS Collaboration~\cite{ATLAS:2013sla}
\be
r_{hVV}=[1.05,1.21],\quad r_{hff}=[0.73,1.07]
\ee
On the other hand in the given range of $v_\Delta$ the signal strengths for the orthogonal eigenstate $S_2$ is very suppressed. For instance for the gluon fusion Higgs production decaying into $VV$ or $ff$ the signal strength is $\propto  r_{S_2 VV (ff)}^2\lesssim 0.1$.  Similarly the pseudoscalar $A$ can decay to fermions with a signal strength $\propto r_{Aff}^2\lesssim 0.1$.

\subsection{The diphoton rate}
\label{diphotonrate}
 
 In this model the extra charged states will contribute to the $S_1\to\gamma\gamma$ decay rate when they propagate in the loop. This rate is dominated in the Standard Model by the propagation of $W$ gauge bosons and top quarks in the loop. The extra contribution from a bosonic or fermionic $Q$-charge sector can be determined from the QED effective potential~\cite{Shifman:1979eb,Carena:2012xa}
 \be
\mathcal L_{\gamma\gamma}=F^2_{\mu\nu}\frac{\alpha}{16\pi}2 \sum_{J,Q}b_{J}^Q\log\det \mathcal M_{J}^Q(X_R),J=0,1/2;\quad X=H_1^0,H_2^0,\phi^0,\psi^0,\chi^0 
\label{formula}
\ee
where $b_{1/2}^{Q_f}=\frac{4}{3} N_c Q_f^2$ for a $Q_f$-charge Dirac fermion, $b_0^{Q_S}=\frac{1}{3} N_c Q_S^2$ for a complex $Q_S$-charge spin-0 boson ($N_c$ being the number of colors of the corresponding field) and where we have subtracted from the determinant in (\ref{formula}) possible zero-modes (e.g.~charged Goldstone bosons).  By expanding $\mathcal L_{\gamma\gamma}$ to linear order in the fields $X_R$ and projecting them into $S_1$ we get for the ratio $r_{S_1\gamma\gamma}$ the general expression
\begin{align}
& \left[A_1(\tau_W)+\frac{4}{3}A_{1/2}(\tau_t)\right]r_{S_1\gamma\gamma}\nonumber\\
&=\left(\cos\alpha_S\cos\theta-\sqrt{\frac{8}{3}}\sin\alpha_S\sin\theta\right)A_1(\tau_W)+\frac{4}{3}\frac{\cos\alpha_S}{\cos\theta} A_{1/2}(\tau_t)\nonumber\\
+&
v_H\left.\left\{\frac{\cos\alpha_S}{\cos\theta}\left( \frac{\partial f}{\partial H_{1R}^0}+\frac{\partial f}{\partial H_{2R}^0}\right)-\sqrt{\frac{2}{3}}\frac{\sin\alpha_S}{\cos\theta}
\left( \frac{\partial f}{\partial \psi_{R}^0}+\frac{\partial f}{\partial \chi_{R}^0}+\frac{\partial f}{\partial \phi_{R}^0}\right)
\right\}\right|_{v_X}
\label{extra}
\end{align}
where $A_1(\tau_W)\simeq -8.3$ and $A_{1/2}(\tau_t)\simeq 1.4$ and
\begin{align}
f(X_R)&=\sum_{Q,J} b_J^Q\log \det\mathcal M_J^Q(X_R)\nonumber\\
&=\frac{4}{3}\log\det \mathcal M_{1/2}^\pm+\frac{16}{3}\log\det \mathcal M_{1/2}^{\pm\pm}
+\frac{1}{6}\log\det \left|\mathcal M_0^\pm\right|^2+\frac{2}{3}\log\det \left|M_0^{\pm\pm}\right|^2
\label{funcionf}
\end{align}
 where we have replaced the minimum conditions (\ref{condiciones}) in the mass matrices $\mathcal M_J^Q$ in (\ref{funcionf}) but \textit{not} the field VEVs so that they depend on the background fields $H_{1R}^0,H_{2R}^0,\phi_R^0,\psi_R^0,\chi_R^0$. In particular it is easy to deduce the contribution from the SM particles $(t,W^\pm)$ in the second line of Eq.~(\ref{funcionf}) from the general expression in the third line of (\ref{extra}) by using the background dependent masses
 \be
 m_t=h_t H_{2R}^0,\quad m_W^2=\frac{1}{2}g^2\left[(H_{1R}^0)^2+(H_{1R}^0)^2+4(\phi_{R}^0)^2+2(\psi_{R}^0)^2+2(\chi_{R}^0)^2\right]\ .
 \ee

\begin{figure}[h!]
\begin{center}
\includegraphics[width=80mm]{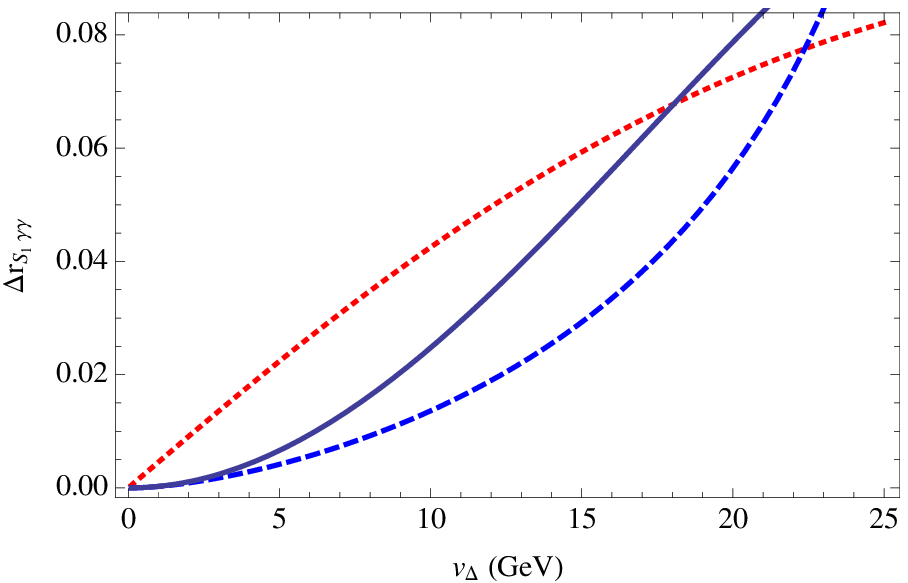}
\includegraphics[width=80mm]{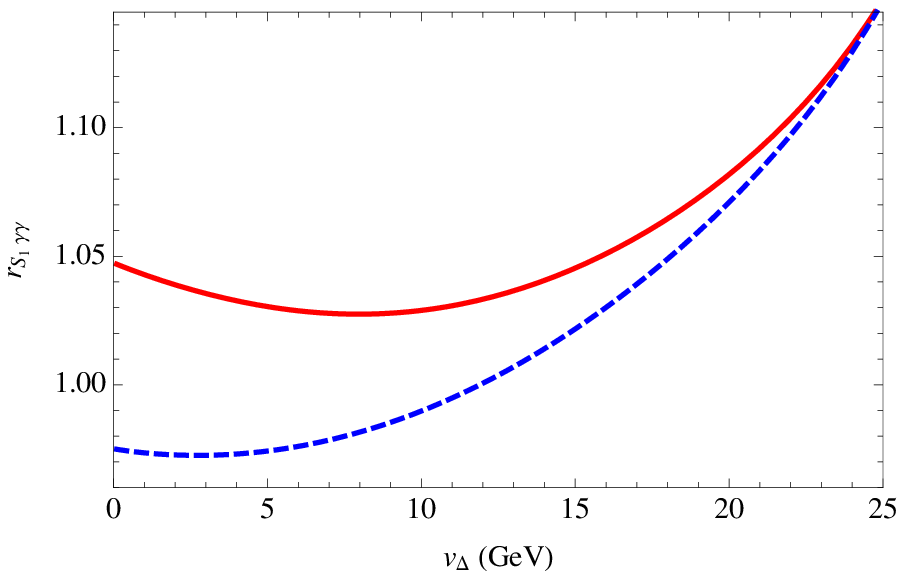}
\end{center}
\caption{\it Left panel: Contribution to $\Delta r_{S_1\gamma\gamma}$ from $W$ and $t$ (solid line), from the doubly charged scalar $F_S^{\pm\pm}$ (dashed line) and from the doubly charged chargino (dotted line) for $M_2=150$ GeV. Right panel: Solid (dashed) line is the plot of $r_{S_1\gamma\gamma}$ for $M_2=$ 150 GeV (300 GeV).}
\label{couplingtophoton}
\end{figure}

As we have doubly charged fields, both in the fermionic and the bosonic sectors, they are expected to dominate the $\gamma\gamma$ production as this one is proportional to $Q^2$. Actually we define the excess in $\gamma\gamma$ with respect to the Standard Model production as 
\be
r_{S_1\gamma\gamma}=1+\Delta r_{S_1\gamma\gamma}
\label{excess}
\ee
 where $\Delta r_{S_1\gamma\gamma}$ is the excess in $r_{S_1\gamma\gamma}$, with respect to the Standard Model contribution, coming from the modified coupling of the Standard Model fields and from the extra charged particles. This excess is plotted in the left panel of Fig.~\ref{couplingtophoton} where the solid line corresponds to the extra contribution from $W$ and $t$ coming from the modified coupling of these particles to the Higgs $S_1$, the dashed line the contribution from the doubly charged scalar $F_S^{\pm\pm}$ which becomes lighter with 
increasing values of $v_\Delta$ (see Fig.~\ref{masas}) and the dotted line the contribution from the doubly charged charginos, where we have taken $M_2=150$ GeV.  The full value of $r_{S_1\gamma\gamma}$ is plotted in the right panel of Fig.~\ref{couplingtophoton} for $M_2=150$ GeV (solid line) and $M_2=300$ GeV (dashed line). 
 
\subsection{Higgs signal strengths} 
\label{gluonfusion}
From the values of $r_{S_1XX}$ determined in the previous
section one can compute the predicted signal strength $\mathcal
R_{S_1 XX}$ of the decay channel $S_1\to XX$, with $X=V,\, f,\,\gamma$:
\be
\mathcal R_{S_1 XX}=\frac{\sigma(pp\to S_1) BR(S_1\to XX)}{\quad\left[\sigma(pp\to h)BR(h\to XX)\right]_{SM}}~.
\ee
In particular for the gluon-fusion (gF), the associated
production with heavy quarks ($S_1tt$), the associated
production with vector bosons ($V S_1$) and the vector boson
fusion (VBF) production processes, one can write
$\mathcal R_{S_1XX}^{(gF)}=\mathcal R_{S_1XX}^{(S_1tt)}=
r_{S_1ff}^2 r_{S_1XX}^2/\mathcal D$ and $\mathcal R_{S_1XX}^{(VBF)}=\mathcal R_{S_1XX}^{(VS_1)}=r_{S_1VV}^2 r_{S_1XX}^2/\mathcal D$ where 
$\mathcal D \simeq
0.74\, r_{S_1ff}^2 + 0.26\, r_{S_1VV}^2$.
\begin{figure}[htb]
\begin{center}
\includegraphics[width=80mm]{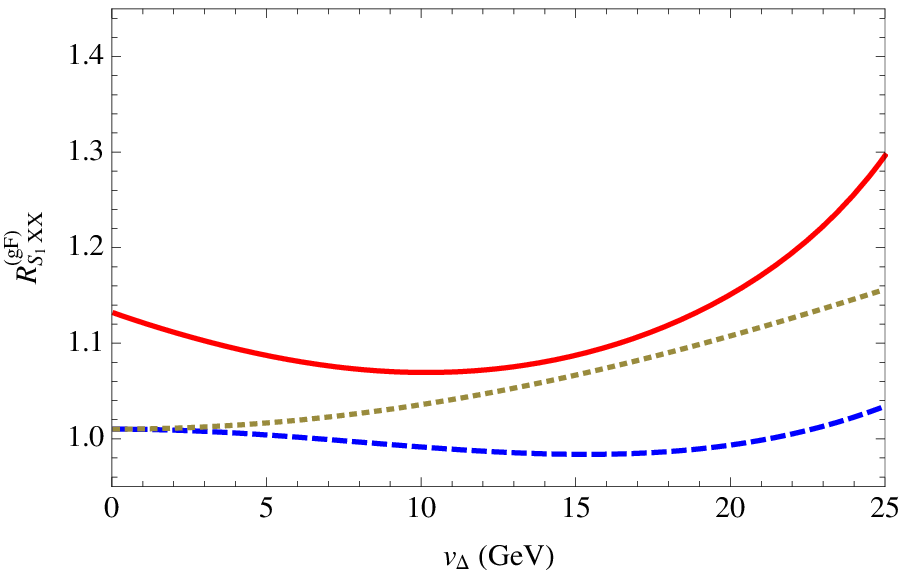}
\includegraphics[width=80mm]{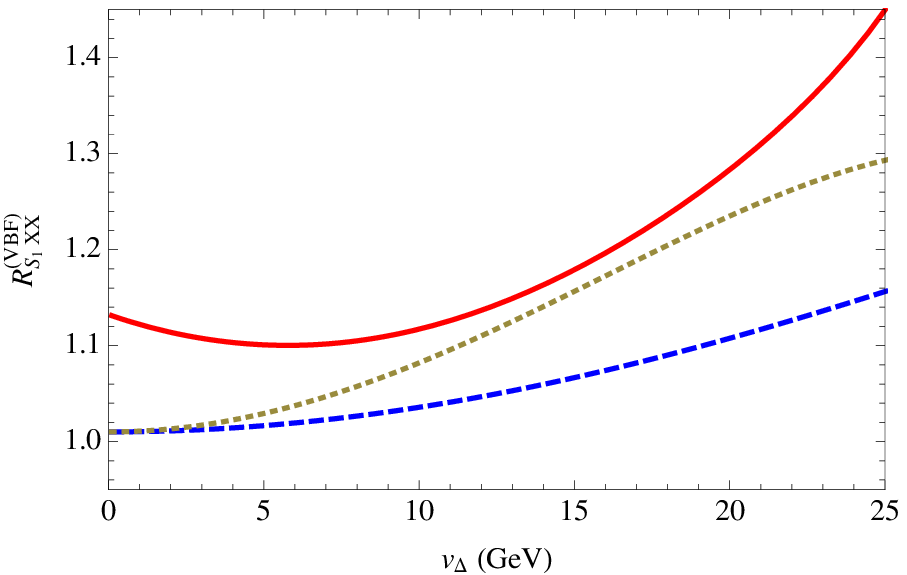}
\end{center}
\caption{\it Left panel: Plot of the gluon fusion Higgs strengths, normalized to the corresponding Standard Model values, for the $\gamma\gamma$ channel (solid line), $bb$ and $\tau\tau$ channels (dashed line) and $WW$ and $ZZ$ channels (dotted line) as a function of $v_\Delta$.  Right panel: The same for weak vector boson fusion Higgs strengths.}
\label{strengths}
\end{figure}
In Fig.~\ref{strengths} we plot $\mathcal R_{S_1XX}^{(gF)}$ (left panel) and $\mathcal R_{S_1XX}^{(VBF)}$ (right panel) for $X=\gamma$ (solid lines), $X=b,\,\tau$ (dashed lines) and $X=W,\, Z$ (dotted lines). The supersymmetric parameters are fixed in Eq.~(\ref{valoresfinales}) and $M_2=150$ GeV. We can see that large values of $v_\Delta$ trigger deviations with the Standard Model expectations although no strong statement can be made at this moment from experimental results. We can just quote the ATLAS best fits for global signal strengths in individual channels~\cite{ATLAS:2013sla}: $\mathcal R_{h\tau\tau}=0.8\pm 0.7$, $\mathcal R_{hWW}=1.0\pm 0.3$ and $\mathcal R_{hZZ}=1.5\pm 0.4$ and $\mathcal R_{h\gamma\gamma}=1.6\pm 0.3$.
 
\section{Breaking custodial symmetry}
\label{breaking}

Up to now we have considered that the Higgs sector respects a global $SU(2)_L\otimes SU(2)_R$ invariance while the vacuum preserves the diagonal (custodial) symmetry subgroup $SU(2)_V$. As the Yukawa and hypercharge couplings explicitly violate the custodial symmetry, radiative corrections, mainly those of the top quark Yukawa coupling, will spoil the custodial invariance of the vacuum. If the custodial vacuum should be considered as a good approximation we should impose some conditions on the fundamental (UV) theory responsible for supersymmetry breaking:
\begin{itemize}
\item
The first condition is that soft supersymmetry breaking generated at some scale $M$ respects the $SU(2)_L\otimes SU(2)_R$ symmetry in the Higgs sector. This means that supersymmetry breaking is generated by effective operators as
\be
\int d^4\theta \frac{X^\dagger X}{M^2} Y^\dagger Y,\quad Y=\bar H,\, \bar\Delta,\, Q,\, L,\, E^c,\,U^c,\,D^c  
\ee
where the spurion field $\langle X\rangle =\theta^2 F$ is responsible for supersymmetry breaking. In other words we will require the same soft mass for scalars in $H_1$ and $H_2$ and similarly for scalars in $\Sigma_0$, $\Sigma_1$ and $\Sigma_{-1}$. This is not a very constraining condition as it is fulfilled e.g.~in minimal SUGRA models~\footnote{Another possibility, that we are not envisaging here, is if soft-breaking terms are not custodial at the scale $M$ and, for some symmetry reason, custodial invariance is an emergent symmetry at the weak scale. Unfortunately we are not aware at this moment of supersymmetry breaking models exhibiting this property.}.
\item
Soft breaking masses run according to the RGE between the scale $M$ and the typical scale where supersymmetric partners decouple which we can identify with $m_{\widetilde Q}$. These radiative corrections spoil the custodial invariance of the minimum through the custodially violating couplings, mainly the top Yukawa coupling. Therefore the second condition is that the breaking induced by RGE between the scales $M$ and $m_{\widetilde Q}$ be consistent with electroweak precision measurement, in particular with the $T$-parameter which measures the failure of custodial invariance. This translates into a \textit{small enough} value of the variable $\log(M/m_{\widetilde Q})$ which is responsible for the RGE running~\footnote{There is another possibility if the soft-breakings are \textit{supersoft}~\cite{Fox:2002bu} in which case there is no RGE running of soft parameters as radiative corrections are finite and there are no large logarithms in the RGE evolution. Notice that in supersoft breaking models the condition $D=0$ required by the custodial symmetry is automatic and, unlike in the MSSM, it has no phenomenological problems for the determination of the Higgs mass.}. We will qualify this second condition in the rest of this section.
 
\end{itemize}

In the theory with $SU(2)_L\otimes SU(2)_R$ invariance in the Higgs sector the minimum equations (\ref{minimo}) at the VEV
\be
v_1=v_H,\quad v_2=v_H\tan\beta,\quad v_\phi=v_\Delta(1+\delta\phi),\quad v_\psi=v_\Delta(1+\delta\psi),\quad
v_\chi=v_\Delta(1+\delta\chi)
\ee
are identically satisfied at the custodial point $\tan\beta=1,\delta\phi=\delta\psi=\delta\chi=0$ and translate into the two conditions, Eqs.~(\ref{condiciones}). Let us consider for simplicity only the leading effect of custodial breaking provided by the top Yukawa coupling~\footnote{Including the  subleading effect of the hypercharge coupling is a trivial generalization of the present case.}. In this case the soft breaking masses of the Higgs fields $H_1$ and $H_2$ are no longer equal
as the top Yukawa coupling induces a negative correction on the mass of $H_2$ as $m_{H_2}^2-m_{H_1}^2=-\delta m_H^2$ with a RGE running controlled by the one-loop $\beta$-function
\be
8\pi^2\beta_{\delta m_H^2}=3h_t^2(m_{H_2}^2+m_Q^2+m_{U^c}^2)+6\chi_2^2(2m_{H_2}^2+m_{\Sigma_{-1}}^2)-6\chi_1^2(2m_{H_1}^2+m_{\Sigma_{1}}^2)
\label{ecuacion}
\ee
where we have neglected the bottom Yukawa coupling, as it is appropriate for small values of $\tan\beta$, and we are using the notation of appendix~\ref{RGE} for the couplings $\chi_{1,2}$ and the masses of $\Sigma_{-1,1}$. A rough approximation (only valid for small enough supersymmetry breaking scales) for the solution to Eq.~(\ref{ecuacion}) can be given in the one-loop approximation since only the first term contributes. In fact the second term is generated at two-loop as the departure from the tree-level relations $\chi_1=\chi_2$ and $m_{\Sigma_{-1}}^2=m_{\Sigma_{1}}^2$ only happens at one-loop. We can then estimate in the one-loop approximation $\delta m_H^2/m_H^2\simeq 3h_t^2(m_{H_2}^2+m_Q^2+m_{U^c}^2)/m_H^2\, \log(M/m_{\widetilde Q})/8\pi^2\simeq (0.1-0.2)\, h_t^2(m_{H_2}^2+m_Q^2+m_{U^c}^2)/m_H^2$ for values $M=10-100$ TeV, while its precise value should depend to a large extent on the particular mechanism of supersymmetry breaking.

Anyway the value generated for $\delta m_H^2$ will perturb the minimum away from the custodial point and will trigger non-zero values of the parameters $\tan\beta-1,\, \delta\phi,\, \delta\psi,\, \delta\chi$. We can analyze this problem numerically by considering small perturbations with respect to the custodial minimum (\ref{condiciones}). We will absorb the modifications produced by $\delta m_H^2$ into a perturbation of $m_H^2$ [$\Delta m_H^2$] as well as on non-zero values of the parameters $\tan\beta-1,\, \delta\phi,\, \delta\psi,\, \delta\chi$: five parameters corresponding to the five minimum equations (\ref{minimo}).

\begin{figure}[htb]
\begin{center}
\includegraphics[width=80mm]{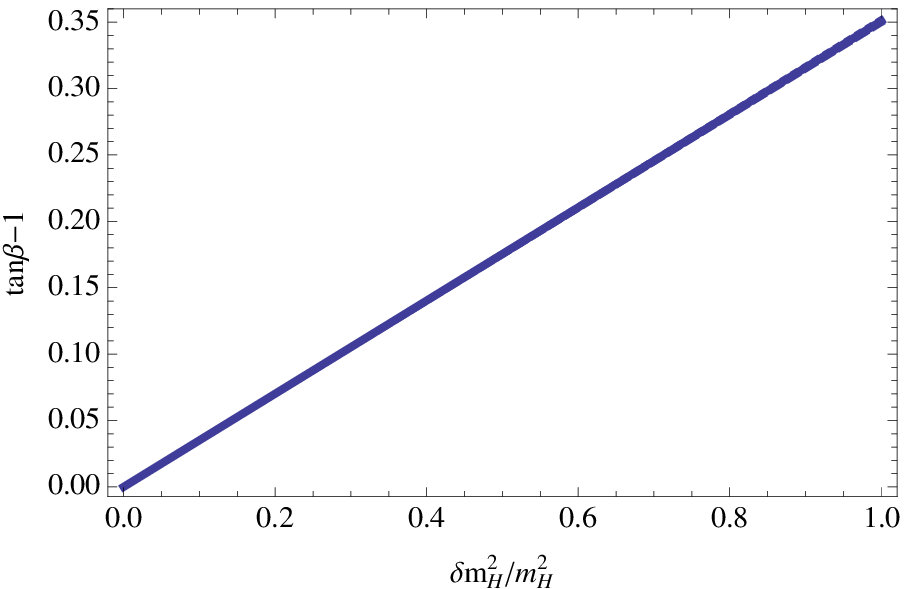}
\includegraphics[width=80mm]{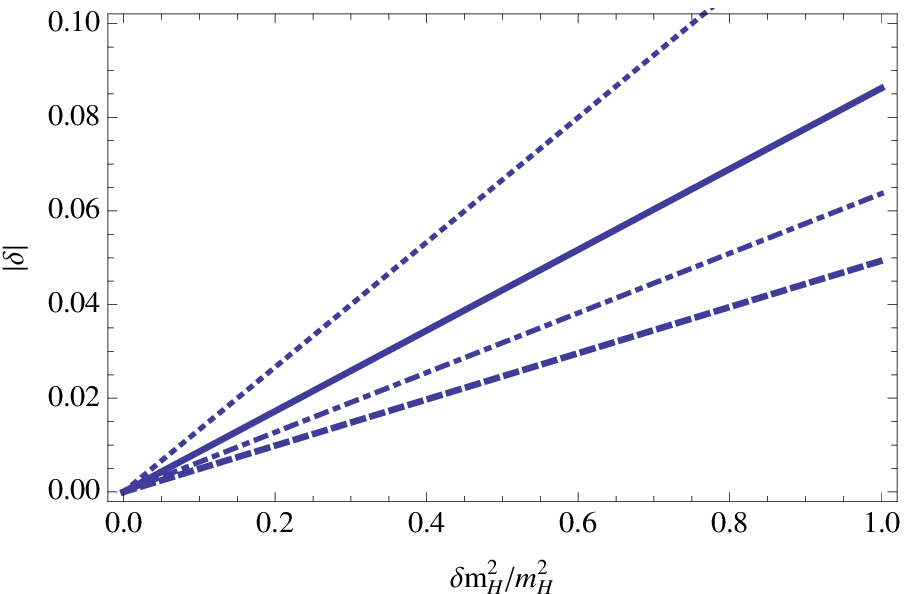}
\end{center}
\caption{\it Left panel: $\tan\beta-1$ as a function of  $\delta m_H^2/m_H^2$ for any value of $v_\Delta$. Right panel: $|\delta|$ as a function of $\delta m_H^2/m_H^2$ for $v_\Delta=$ 5 GeV (dotted line), 10 GeV (solid line), 15 GeV (dot dashed line) and 20 GeV (dashed line). Parameters are fixed as in (\ref{valoresfinales}) and $m_{S_1}=126$ GeV.}
\label{perturbations}
\end{figure}
From the equation $\partial V/\partial H_{1R}^0=0$ we can trade the value of $\Delta m_H^2$ in terms of other variables as 
\be
\Delta m_H^2 =\left[ m_3^2-\left(2\lambda^2-\frac{G^2}{2}\right)\right](\tan\beta-1)+\mathcal O(v_\Delta)
\label{Deltafinal}
\ee
where we have made a series expansion in powers of $v_\Delta$. From (\ref{Deltafinal}) we see that a non-zero value of $tan\beta-1$ [as well as non-zero values of $(\delta\phi,\, \delta\psi,\, \delta\chi)$ to subleading order] will trigger the perturbation of $m_H^2$,  $\Delta m_H^2$. Moreover from the other four equations non-zero values of $(\tan\beta-1,\,\delta\phi,\, \delta\psi,\, \delta\chi)$ will be triggered by the non-zero value of $\delta m_H^2$. In particular the non-zero value of $(\delta\phi,\, \delta\psi,\, \delta\chi)$ will produce a non-zero value of the $T$ observable as
\be
\alpha T=4 \delta\frac{v_\Delta^2}{v^2},\quad \delta=2\delta\phi- \delta\psi-\delta\chi
\ee
which is constrained to the region $-0.09<T<0.23$ at the 95\% CL~\cite{Beringer:1900zz}.

We will linearize the four equations $\partial V/\partial H_{2R}^0=\partial V/\partial \phi_{R}^0=\partial V/\partial \psi_{R}^0=\partial V/\partial \chi_{R}^0=0$ in the variables $(\tan\beta-1,\,\delta\phi,\, \delta\psi,\, \delta\chi)$ and numerically check a posteriori the self consistency of this approximation. The relevant results are plotted in Fig.~\ref{perturbations} where we show $\tan\beta-1$ (left panel)  and $|\delta|$ (right panel) as functions of $\delta m_H^2/m_H^2$ and where we have fixed the values of supersymmetric parameters as in (\ref{valoresfinales}) and $m_{S_1}=126$ GeV.
\begin{figure}[htb]
\begin{center}
\includegraphics[width=100mm]{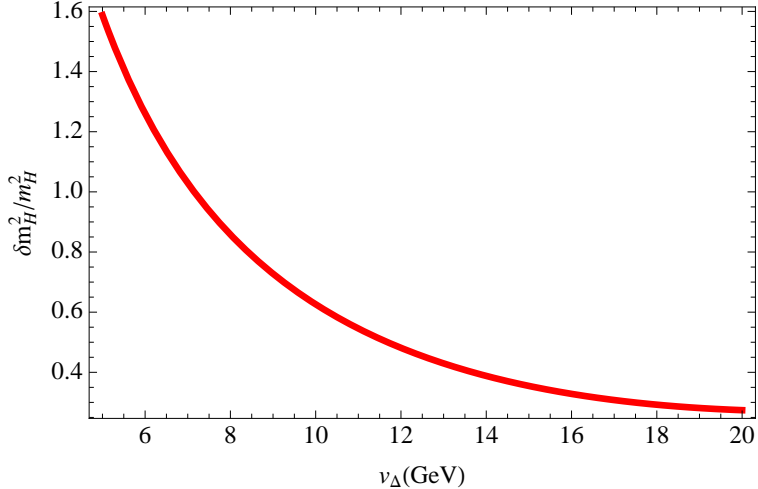}
\end{center}
\caption{\it 95\% CL allowed region in the $(\delta m_H^2/m_H^2,v_\Delta)$ plane  (below, and on the left, of the solid line). Parameters are fixed as in (\ref{valoresfinales}) and $m_{S_1}=126$. GeV.}
\label{T}
\end{figure}
From the left panel of Fig.~\ref{perturbations} we see that the departure of $\tan\beta$ from one is less than 30\% while the size of $\delta$ depart from zero by less than 0.1 in all cases. As a consequence the allowed region in the $(\delta m_H^2/m_H^2,v_\Delta)$ plane is shown in the right panel of Fig.~\ref{T} at the 95\% CL.

Notice that, as we already pointed out, the value of the parameter $\delta m_H^2/m_H^2$ is determined to a large extent by the supersymmetry breaking scale $M$ and by the squark spectrum at the scale $M$. So the results from Figs.~\ref{perturbations} and \ref{T} imply in general a low value of $M$: the larger $v_\Delta$ the lower $M$. In the absence of any particular model of supersymmetry breaking we can roughly fix the interval as $M\sim$10 TeV (for the large values of $v_\Delta$)-100 TeV (for the small values of $v_\Delta$).

\section{Conclusion}
\label{conclusion}

In this paper we have proposed a supersymmetric model with $Y=(0,\pm 1)$ triplets whose Higgs sector exhibits a global $SU(2)_L\otimes SU(2)_R$ symmetry which is spontaneously broken to the diagonal (custodial) $SU(2)_V$ in the theory vacuum. Of course supersymmetry is motivated by the solution to the hierarchy problem. The model consists in a supersymmetrization of the (non-supersymmetric) model proposed long ago by Georgi and Machacek in Ref.~\cite{Georgi:1985nv} where the tree-level $\rho$ parameter is kept to unity by the remaining custodial symmetry even if the neutral components of Higgs triplets acquire a non-vanishing VEV $v_\Delta$. Moreover we expect that supersymmetry will take care of the quadratic sensitivity exhibited by radiative corrections to the $\rho$ parameter in the original GM model~\cite{Gunion:1990dt}. In spite of having three supersymmetric triplets on top of the MSSM superfields the bosonic sector of the model has a very simple structure as a consequence of the underlying $SU(2)_V$ symmetry: scalars are classified into singlets, triplets and fiveplets with degenerate masses.

The conditions for electroweak symmetry breaking at the custodial minimum are easily satisfied. Although we have not investigated other possible local minima where custodial symmetry might be explicitly broken we expect such a non-custodial minimum to be energetically disfavored with respect to the custodial one. Neither have we tried to scan over the parameter space of the model. Instead we have chosen generic values of the supersymmetric parameters and adjusted, for every value of $v_\Delta$, the custodial superpotential coupling $\lambda=\lambda(v_\Delta)$ of triplets to the MSSM Higgs sector such as the mass of the SM-like scalar reproduces the experimentally observed value of the Higgs boson mass at LHC $\sim$ 126 GeV. We have found, unlike in the MSSM, two different decoupling regimes: \textbf{i)} The limit where $m_\Delta\to\infty$ in which case the triplet scalars become infinitely heavy and are decoupled from the doublet scalars. From the minimization equations this limit amounts to $v_\Delta\to 0$. Unlike in the custodially unprotected models this limit is not compelling in our model as the tree-level value of the $\rho$ parameter is one; \textbf{ii)} The limit where $m_3^2\to\infty$, where $m_3^2$ is the soft breaking coefficient of the Lagrangian term $H_1\cdot H_2$. This decoupling regime is similar to the MSSM one. Actually in the whole decoupling regime the only surviving light state is the SM-like Higgs as expected. On the other hand, as lots of LHC phenomenological studies are performed for the non-supersymmetric GM model~\cite{GMstudies} it should be interesting to recover the latter from our supersymmetric model~\cite{preparation}.

The couplings of the SM-like boson to vector bosons (including $\gamma\gamma$) and fermions are computed along the trajectory $\lambda(v_\Delta)$ and their value as well as the Higgs signal strengths for the different Higgs production mechanisms (gluon-fusion, vector boson fusion and associated production with tops and vector bosons) is fully consistent with the ATLAS and CMS results within experimental uncertainties. We certainly expect that at LHC14 the experimental results on the different Higgs signal strengths will constrain the parameter range of our model. 

An issue which we have only partially considered in this paper is that of the transmission of custodial breaking by radiative corrections mainly driven by the custodial violating top Yukawa and hypercharge couplings. In particular we have considered the splitting in the soft masses of the MSSM Higgses, $H_1$ and $H_2$, triggered by the leading effect of the top Yukawa coupling and transmitted by the renormalization group equations. Under the assumption of a custodial mechanism of supersymmetry breaking this leads to the requirement of low-scale supersymmetry breaking. Of course quantifying the smallness of the supersymmetry breaking scale $M$ is to a large extent model dependent. In the absence of a particular mechanism of supersymmetry breaking we can just postulate values $M\sim 10-100$ TeV. A systematic loop analysis from electroweak observables in this model is compulsory and should be done in order to constrain regions in supersymmetric parameter space~\cite{preparation}. 
The model also has new (bosonic and fermonic) singly and doubly charged states with a rich phenomenology which should yield typical signatures at the LHC, which we have not tried to explore in this paper but that would be worth investigating in the future~\cite{preparation}.

Finally we would like to mention that this model, similarly to (supersymmetric or non-supersymmetric) type-II seesaw models~\cite{Gunion:1989ci,Hambye:2000ui,Ma:2000xh,Chun:2003ej,1,Dev:2013ff}, can accommodate a renormalizable contribution to the Majorana neutrino masses through a $\Delta L=2$, custodial-violating, superpotential term like
$
W_\nu=h^{ij}_\nu L_i\Sigma_1 L_j
\label{ultimo}
$,
which yields a neutrino Majorana mass matrix $\mathcal M_\nu^{ij}=h^{ij}_\nu v_\Delta$. For the values of $v_\Delta$ used throughout this paper, the Yukawa couplings $h^{ij}_\nu$ are tiny and their breaking of the custodial invariance negligible.

\section*{\sc Acknowledgments}
Work supported by the Spanish Consolider-Ingenio 2010 Programme CPAN (CSD2007-00042) and by CICYT-FEDER-FPA2011-25948. We thank S. Gori and R.~Vega-Morales for a careful reading of the manuscript and R.~Vega-Morales for useful comments about the quadratic sensitivity of the $\rho$ parameter in the Georgi-Machacek model.

\section*{Appendix}
\appendix
\section{The model potential}
 \label{modelpotential} 
 The superpotential (\ref{W0}) in component fields is given by
 \begin{align}
W_0&=\lambda\left[H_1^0\phi^0H_2^0+H_2^0\chi^0 H_2^0+H_1^0 \psi^0 H_1^0+H_1^-\phi^0H_2^+-H_2^+\chi^{--}H_2^+-H_1^- \psi^{++}H_1^- 
 \right. \nonumber\\
 &+\left.\sqrt{2}( H_1^-H_2^0\phi^++H_2^+H_2^0\chi^--H_1^-H_1^0\psi^+-H_2^+H_1^0\phi^- )\right]\nonumber\\
 &+\lambda_3\left[-\phi^0\chi^0\psi^0+\psi^{++}\phi^0\chi^{--}+\phi^+\psi^0\chi^-+\phi^-\chi^0\psi^+-\phi^+\chi^{--}\psi^+-\phi^-\psi^{++}\chi^-
 \right]\nonumber\\
 &+\mu_\Delta\left[\frac{1}{2}\phi^{0}\phi^0+\psi^0\chi^0+\phi^+\phi^-+\psi^+\chi^-+\psi^{++}\chi^{--}
 \right]+\mu\left[H_1^-H_2^+-H_1^0H_2^0   \right]
\end{align}
 and correspondingly the F-term potential is given by
 \begin{eqnarray}
 V_F&=&\left| \lambda\left(\phi^0 H_2^0+2 H_1^0\psi^0-\sqrt{2}H_1^-\psi^+-\sqrt{2}H_2^+\phi^-\right)-\mu H_2^0    
 \right|^2 \nonumber\\
 &+&\left|     \lambda\left(\phi^0 H_1^0+2 H_2^0\chi^0+\sqrt{2}H_1^-\phi^+ +\sqrt{2}H_2^+\chi^-\right)-\mu H_1^0
 \right|^2 \nonumber\\
 &+&\left|  \lambda(H_1^0 H_2^0+H_1^-H_2^+)-\lambda_3(\chi^0\psi^0-\psi^{++}\chi^{--})+\mu_ \Delta \phi^0   
 \right|^2 \nonumber\\
 &+&\left| \lambda H_1^0 H_1^0-\lambda_3(\phi^0\chi^0-\phi^+\chi^-)+\mu_\Delta \chi^0
 \right|^2 \nonumber\\
  &+&\left| \lambda H_2^0 H_2^0-\lambda_3(\phi^0\psi^0-\phi^-\psi^+)+\mu_\Delta \psi^0
 \right|^2 \nonumber\\
 &+&\left|  \lambda(\phi^0 H_2^+-2\psi^{++}H_1^- +\sqrt{2} H_2^0\phi^+-\sqrt{2}H_1^0\psi^+)+\mu H_2^+   
 \right|^2 \nonumber\\
 &+&\left| \lambda(\phi^0 H_1^- -2\chi^{--}H_2^+ +\sqrt{2} H_2^0\chi^- -\sqrt{2}H_1^0\phi^-)+\mu H_1^-     
 \right|^2 \nonumber\\
 &+&\left|  \sqrt{2} \lambda H_1^- H_2^0+\lambda_3(\psi^0\chi^- - \chi^{--}\psi^+)+\mu_\Delta \phi^-   
 \right|^2 \nonumber\\
 &+&\left|     -\sqrt{2} \lambda H_1^0 H_2^+ +\lambda_3(\chi^0\psi^+ - \psi^{++}\chi^-)+\mu_\Delta \phi^+   
 \right|^2 \nonumber\\
 &+&\left| -\sqrt{2} \lambda H_1^- H_1^0+\lambda_3(\phi^-\chi^0-\phi^+\chi^{--})+\mu_\Delta\chi^-  
 \right|^2 \nonumber\\
 &+&\left|  \sqrt{2}\lambda H_2^0H_2^+ +\lambda_3(\phi^+\psi^0-\phi^-\psi^{++})+\mu_\Delta \psi^+  
 \right|^2 \nonumber\\
 &+&\left| -\lambda H_1^- H_1^-+\lambda_3(\phi^0\chi^{--}-\phi^-\chi^-)+\mu_\Delta\chi^{--}    
 \right|^2 \nonumber\\
 &+&\left|  -\lambda H_2^+ H_2^+ +\lambda_3(\phi^0\psi^{++}-\phi^+\psi^+)+\mu_\Delta\psi^{++}    
 \right|^2 
 \end{eqnarray} 
  
  The soft breaking potential (\ref{Vsoft}) can be written as
  \begin{align}
  V_{\rm soft}&=m_H^2\left(|H_1^0|^2+ |H_2^0|^2+|H_1^-|^2+|H_2^+|^2
  \right)\nonumber\\
  &+m_\Delta^2\left(|\phi^0|^2+|\phi^-|^2+|\phi^+|^2+|\psi^0|^2+|\psi^+|^2+|\psi^{++}|^2+|\chi^0|^2+|\chi^-|^2+|\chi^{--}|^2 \right)\nonumber\\
  &+\left\{m_3^2(H_1^- H_2^+-H_1^0 H_2^0)+B_\Delta(\phi^0\phi^0/2+\psi^0\chi^0+\phi^+\phi^-+\psi^+\chi^-+\psi^{++}\chi^{--}  )\right.\nonumber\\
  & + A_{\lambda_3}\left( -\phi^0\chi^0\psi^0+\phi^0\psi^{++}\chi^{--}+\phi^+\psi^0\chi^-+\phi^-\chi^0\psi^+-\phi^+\chi^{--}\psi^+-\phi^-\psi^{++}\chi^-
   \right)\nonumber\\
   &+A_\lambda\left(H_1^0\phi^0H_2^0+H_2^0\chi^0 H_2^0+H_1^0\psi^0 H_1^0+H_1^-\phi^0 H_2^+-H_2^+ \chi^{--}H_2^+-H_1^-\psi^{++}H_1^-
      \right.\nonumber\\
   &+\left.\sqrt{2}\left. \left[ H_1^-\phi^+ H_2^0-H_1^0\phi^- H_2^++ H_2^0\chi^- H_2^+-H_1^-\psi^+ H_1^0      \right] \right)+h.c.\right\}
  \end{align}
 
 Finally the $D$-terms are
 \begin{align}
 D^1-iD^ 2&=-g\left\{H_1^{0*} H_1^- +H_2^{+*} H_2^0\right.\nonumber\\
 &+\sqrt{2}\left.\left[\phi^{0*}\phi^- -\phi^0\phi^{+*}+\psi^{+*}\psi^{0}-\psi^+\psi^{++*}+\chi^{-*}\chi^{--} -\chi^{-}\chi^{0*}\right]\right\}
 \nonumber\\
 D^1+iD^2&=(D^1-iD^2)^*\nonumber\\
  D^3&=-\frac{g}{2}\left\{|H_1^0|^2-|H_2^0|^2-|H_2^+|^2-|H_1^-|^2\right.\nonumber\\
 &+\left.2\left[|\chi^0|^2-|\psi^0|^2+|\phi^+|^2-|\phi^-|^2+|\psi^{++}|^2-|\chi^{--}|^2\right]
 \right\}
 \nonumber\\
 D_Y&=-\frac{g'}{2}\left\{|H_2^0|^2-|H_1^0|^2+|H_2^+|^2-|H_1^-|^2\right.\nonumber\\
 &+\left.2\left[|\psi^0|^2-|\chi^0|^2+|\psi^{++}|^2+|\psi^{+}|^2-|\chi^{--}|^2-|\chi^-|^2\right]
 \right\}\end{align}
 and the D-potential is given by
 \begin{align}
 V_D&=\frac{g^2}{8} \left\{|H_1^0|^2-|H_2^0|^2+|H_2^+|^2-|H_1^-|^2+2|\chi^0|^2-2|\psi^0|^2\right.\nonumber\\
&+\left. 2|\phi^+|^2-2|\phi^-|^2+2|\psi^{++}|^2-2|\chi^{--}|^2\right\}^2 \nonumber\\
&+\frac{g^2}{2}\left|H_1^0 H_1^{-\,*}+H_2^{0\,*}H_2^+\right.\nonumber\\
&+\left.\sqrt{2}\left[\phi^{0}\phi^{-*} -\phi^0\phi^{+}+\psi^{+}\psi^{0*}-\psi^{+*}\psi^{++}+\chi^{-}\chi^{--*} -\chi^{-*}\chi^{0}\right]
\right|^2\nonumber\\
&+\frac{g^{\prime 2}}{8} \left\{|H_2^0|^2-|H_1^0|^2+|H_2^+|^2-|H_1^-|^2+2|\psi^0|^2-2|\chi^0|^2\right.\nonumber\\
&+\left. 2|\psi^+|^2-2|\chi^-|^2+2|\psi^{++}|^2-2|\chi^{--}|^2\right\}^2 
 \end{align} 
  
  \section{The custodial $SU(2)_V$}
  \label{su2}
A finite $SU(2)_L\otimes SU(2)_R$ transformation acting on the Higgs antidoublet 
\be
\bar{H} \rightarrow   \mathcal{U}_L\otimes\mathcal{\bar{U}}_R\,\,\bar{H}
\ee
leads to the usual transformation
\be
\bar{H}^T \rightarrow \mathcal{U}_L\bar{H}^T \mathcal{U}_R^\dagger \, .
\ee
The condition for this multiplet to break $SU(2)_L\otimes SU(2)_R$ to $SU(2)_V$ is,
\be
\langle \bar{H}^T \rangle  =  \mathcal{U}_L\langle \bar{H}^T \rangle \mathcal{U}_R^\dagger \, ,
\ee
\noindent in detail,
\be
\begin{pmatrix} v_1 & 0 \\ 0 & v_2 \end{pmatrix} =\exp{\{i\theta_L^a \sigma_a/2\}}\begin{pmatrix} v_1 & 0 \\ 0 & v_2 \end{pmatrix} \exp{\{-i\theta_R^a \sigma_a/2\}} \, .
\ee
If we make $\theta_L=\theta_R$ and $v_1=v_2\equiv v_H$ the vacuum leaves unbroken the vectorial subgroup $SU(2)_{L+R}=SU(2)_V$ i.e. the custodial symmetry. \\

The same game can be played for the antitriplet. Now $SU(2)$ transformations will act on both sides, as
\be
\bar{\Delta}  \rightarrow  \bar{U}_R\otimes U_L\,\bar{\Delta}\, U_L^\dagger\otimes \bar{U}_R^\dagger \, ,
\ee
\noindent where $U_L= \exp{\{i\theta_L^a \sigma_a/2\}}$ and $\bar U_R= \exp{\{i\theta_R^a \bar \sigma_a/2\}}$ \footnote{Note that for the complex conjugate of any $SU(N)$ representation, the generators are $T_{\bf{\bar{n}}}^a=-(T_{\bf{n}}^a)^*$.}. The condition is now,
\be
\langle\bar{\Delta}\rangle  =\bar{U}_R \begin{pmatrix} 
- U_L \frac{1}{\sqrt{2}}\langle\Sigma_0\rangle U_L^\dagger & - U_L\langle\Sigma_{-1}\rangle U_L^\dagger \\
-U_L\langle\Sigma_1\rangle U_L^\dagger & U_L\frac{1}{\sqrt{2}}\langle\Sigma_0\rangle U_L^\dagger
\end{pmatrix} \bar{U}_R^\dagger \, . 
\ee
This relation is preserved (the unbroken subgroup is the vectorial one) if $v_\chi=v_\psi=v_\phi \equiv v_\Delta$ and $\theta_L=\theta_R$. \\

One could also use the vector representation, $\bar\Delta^T=(\Sigma_{-1},\Sigma_0,\Sigma_1)$. In this case,
\be
\bar{\Delta}^T  \rightarrow  \mathcal{U}_L\bar{\Delta}^T\mathcal{U}_R^\dagger \, ,
\ee
\noindent with $\mathcal{U}_L=\exp{\{i\theta_L^a t_a\}}$ and $\mathcal{U}_R=\exp{\{i\theta_R^a t_a\}}$, where $t_a$ are the generators  of SU(2) in the representation \textbf{3}
\begin{equation}
t_a=\{t_1,t_2,t_3\}=\left\{\frac{1}{\sqrt{2}}\begin{pmatrix} 
0 & 1 & 0 \\
1 & 0 & 1 \\
0 & 1 & 0
\end{pmatrix},\,\frac{1}{\sqrt{2}}\begin{pmatrix} 
0 & -i & 0 \\
i & 0  & -i \\
0 & i & 0
\end{pmatrix},\,\begin{pmatrix} 
1 & 0 & 0 \\
0 & 0 & 0 \\
0 & 0 & -1 
\end{pmatrix} \right\}
\end{equation}
Using them this calculation leads to the same results. 

Another check to see whether the new particle content preserves the custodial symmetry is to compute the tree-level contributions to the $\rho$ parameter. The only extra contribution will come from,
\be
\frac{1}{2}\mathrm{Tr}[(D_\mu\bar{\Delta})^\dagger(D^\mu\bar{\Delta})] \, .
\ee
Note that the definition of the covariant derivative changes here, as we identify $Y=T_{3R}$, i.e.
\be
D_\mu\bar{\Delta}  =  \partial_\mu\bar{\Delta}+igW_\mu^at^a\bar{\Delta}-ig^\prime\bar{\Delta} B_\mu Y t^3 \, .
\ee
Using this definition we compute the contributions coming from $\bar\Delta$, 
\begin{eqnarray}
\frac{1}{2}\mathrm{Tr}[(D_\mu\bar{\Delta})^\dagger(D^\mu\bar{\Delta})]  & \supset &g^2\left[ v_\chi^2\,W_\mu^+W^{\mu-}+\frac{v_\chi^2}{\cos^2\theta_W}Z_\mu Z^\mu\right. \nonumber \\ 
& + & v_\phi^2\,W_\mu^+W^{\mu -}+v_\phi^2\,W_\mu^-W^{\mu+} \\ & + &
\left.v_\psi^2\,W_\mu^-W^{\mu+}+\frac{v_\psi^2}{\cos^2\theta_W}Z_\mu Z^\mu\right] \nonumber
\end{eqnarray}
If we choose the custodially preserving direction of the vacuum ($v_\phi=v_\chi=v_\psi\equiv v_\Delta$ and $v_1=v_2\equiv v_H$) the gauge boson masses,
\be
m_W^2=\cos^2\theta_W m_Z^2=\frac{g^2(2v_H^2+8v_\Delta^2)}{2} \, ,
\ee
this keeps $\rho=1$ and fixes the value of the electroweak VEV, 
\be
v^2=2v_H^2+8v_\Delta^2 \, .
\ee

As it happens in the MSSM (or in any model with only doublets in the Higgs sector), where the $\rho$ parameter is protected at tree level for any value of $v_1$ and $v_2$, there is a more general direction for the VEV's that preserves $\rho=1$. In fact if $v_\phi^2=\frac{1}{2}(v_\psi^2+v_\chi^2)$, $\rho=1$ is ensured at tree level and $v^2=v^2_{MSSM}+4(v_\psi^2+v_\chi^2)=v^2_{MSSM}+8v_\phi^2$. However the direction that keeps the custodial symmetry safe is stronger than $\rho=1$ at tree level, as it also prevents corrections to the $\rho$ parameter at loop level from superpotential couplings. 

\section{RGE for Yukawa couplings}
\label{RGE}

The superpotential (\ref{W0}) can  be written in component superfields as
\begin{align}
	W_0 & =-\lambda H_1\cdot\Sigma_1 H_1+\lambda H_2 \cdot \Sigma_{-1} H_2+\sqrt{2}\lambda H_1\cdot \Sigma_0 H_2+  \sqrt{2}\lambda_3\tr\Sigma_0\Sigma_{-1}\Sigma_1  \\
	& + \frac{\mu}{2}(H_1\cdot H_2-H_2\cdot H_1)+\mu_\Delta\tr(\Sigma_0^2+\Sigma_1\Sigma_{-1}) \nonumber \, .
\end{align}
However in our model, when computing the beta functions for the Yukawa parameters appearing in the superpotential, one has to keep in mind that the breaking induced by $g^\prime$ and the top Yukawa coupling will split the custodially preserving coefficients written in (\ref{W0}). Because of this we should write the general superpotential in terms of $SU(2)_L$ multiplets where custodial invariance is not implemented. In particular the trilinear superpotential involving the Yukawa parameters, using  the notation previously used in Ref.~\cite{cuatro}, reads as
\begin{align}
W_{\textrm{3}} & =\chi_1 H_1\cdot\Sigma_1 H_1+\chi_2 H_2 \cdot \Sigma_{-1} H_2+\sqrt{2}\lambda_2 H_1\cdot \Sigma_0 H_2\nonumber\\ 
 &+ \sqrt{2} \lambda_0\tr\Sigma_0\Sigma_{-1}\Sigma_1+  h_tQ\cdot H_2 U^c \, .
\end{align}

With this notation the RGE's for the Yukawa parameters appearing in the superpotential are~\footnote{Note that we do not consider the contribution coming from $h_b$, a good approximation in a small $\tan\beta$ regime.},
\begin{eqnarray}
 4\pi^2\frac{d}{dt}\lambda_2^2 & = & \left(-\frac{7}{2}g^2-\frac{1}{2}g^{\prime\, 2} + \frac{3}{2}h_t^2+4\lambda_2^2+3(\chi_1^2+\chi_2^2)+\lambda_0^2 \right) \lambda_2^2 \\
  4\pi^2\frac{d}{dt}\chi_1^2 & = & \left(-\frac{7}{2}g^2-\frac{3}{2}g^{\prime\, 2}+3\lambda_2^2+7\chi_1^2+\lambda_0^2\right)\chi_1^2 \\
  4\pi^2\frac{d}{dt}\chi_2^2 & = & \left(-\frac{7}{2}g^2-\frac{3}{2}g^{\prime\, 2}+3h_t^2+3\lambda_2^2+7\chi_2^2+\lambda_0^2\right)\chi_2^2 \\
  4\pi^2\frac{d}{dt}\lambda_0^2 & = & \left(-6g^2-2g^{\prime\, 2}+\chi_1^2+\chi_2^2+\lambda_2^2+3\lambda_0^2\right)\lambda_0^2 \\
  8\pi^2 \frac{d}{dt}h_t & = & \left(-\frac{3}{2}g^2-\frac{13}{18}g^{\prime\, 2}-\frac{8}{3}g_s^2+3h_t^2+\frac{3}{2}\lambda_2^2+3\chi_2^2\right) h_t \\
  16\pi^2 \frac{d}{dt} g & = & 7g^3 \\
  16\pi^2 \frac{d}{dt} g^\prime & = & 17g^{\prime\, 3} \\ 
  16\pi^2 \frac{d}{dt} g_s & = & -3g_s^3 
\end{eqnarray}
with initial conditions fulfilling the custodial symmetry at the weak scale
\be
\lambda_2^2(m_t)=\lambda^2,\quad \chi_1^2(m_t)=\chi_2^2(m_t)=\lambda^2, \quad \lambda_0^2(m_t)=\lambda_3^2, \quad h_t(m_t)=\frac{\sqrt{2}m_t}{\sqrt{v^2-4v_\Delta^2}}.
\ee

As we can see from the RGE's custodial symmetry is spoiled away of the electroweak scale by the hypercharge and top Yukawa couplings. Moreover the theory becomes non-perturbative at some scale which depends of course on the initial conditions for the different parameters. For the chosen values of the supersymmetric parameters in (\ref{valores}) the location of the corresponding Landau pole is plotted in Fig.~\ref{lambda} (right panel). The theory requires then UV completion at scales below the location of the Landau pole and in particular the scale of supersymmetry breaking should also be below it, which leads to a theory of low-scale supersymmetry breaking.

\end{document}